\newif\ifnote			
\newif\ifsupplementary	
\newif\ifnature			
\newif\ifnaturesubmit
\newif\ifincludemain

\notetrue
\includemaintrue
\supplementarytrue
\naturetrue

\newcommand{\he}[1]{$^{#1}$He}

\newcommand{\singlequote}[1]{`\textit{#1}'}

\ifsupplementary
	
\else
	
\fi

\ifnote
	
\else
	
\fi

\ifnature
	\newcommand{\eqref}[1]{(\ref{#1})}
	
	\newcommand{\mySection}[1]{\section*{#1}}
	
	\documentclass{nature}
	
	\author{Robert M Brady$^{1}$, Edward T Samulski$^2$  \& David H P Turban$ ^3 $}

\else
	\newcommand{\mySection}[1]{\textbf{#1}}

	\documentclass[12pt]{article}
	\usepackage[margin=20mm]{geometry}
	\usepackage{amssymb,amsmath}
	\usepackage[super]{cite} 

	\author{Robert M Brady$ ^* $ and Edward T Samulski$ ^\dagger $ \\
	\footnotesize * University of Cambridge Computer laboratory, JJ Thomson Avenue, Cambridge CB3 0FD, UK \\
	\footnotesize $ \dagger  $ Department of Chemistry CB\#3290, University of North Carolina at Chapel Hill, North Carolina 27599, USA }

\fi

\usepackage{graphicx}
\usepackage{xcolor}
\usepackage{enumerate}
\usepackage{amssymb}
\usepackage{hyperref}
\pdfoutput=1

\title{Evidence that rotons in helium II are interstitial atoms}

\begin{document}

\maketitle


\ifnature
		\begin{affiliations}
			\item University of Cambridge Computer laboratory, JJ Thomson Avenue, Cambridge CB3 0FD, UK 
			\item Department of Chemistry CB\#3290, University of North Carolina at Chapel Hill, North Carolina 27599, USA
			\item University of Cambridge Cavendish Laboratory, JJ Thomson Ave, Cambridge CB3 0HE, UK
	\end{affiliations}

\ifincludemain

\begin{abstract}

Superfluid helium II
contains excitations known as rotons.
Their properties have been studied experimentally
for more than 70 years but their structure
is not fully understood.
Feynman's 1954 description, involving rotating flow patterns, 
does not fully explain later experimental data.
Here we identify volumetric, thermodynamic, colloidal, excitation, x-ray and neutron scattering evidence
that rotons are composed of interstitial helium atoms.
We show in particular that they 
have the same mass,
effective mass and activation energy 
within experimental accuracy.
They readily move through the substrate, and
couple through lattice vibrations to produce quantized,
loss-free flow which corresponds to the observed superflow.
Our observations revive London's 1936 conclusion that helium II
has a relatively open crystal-like lattice
with enough free volume for atoms to move relative to one another, and reconcile it with London's 1938
description of a quantum fluid.

\end{abstract}


\ifsupplementary
{\color{blue} Including supplementary material on page \pageref*{sec:supplementary-material}.
	
}
\fi

\ifnature
\else
	\setlength{\parskip}{3.5mm plus1mm minus1mm}
\fi

\mySection{Condensed phases of helium}

Every known element except helium 
has a triple point
where solid, liquid and vapour coexist\cite{young1975phase}.
\he4lium (figure \ref{fig:phase-diagram-he4}) has 
a flowing phase, helium II,
where a solid would be expected\cite{donnelly1998observed,brooks1977calculated,hoffer1976thermodynamic}.

\begin{figure}[htb]
	\centering
	
	\includegraphics[width=.85\linewidth]{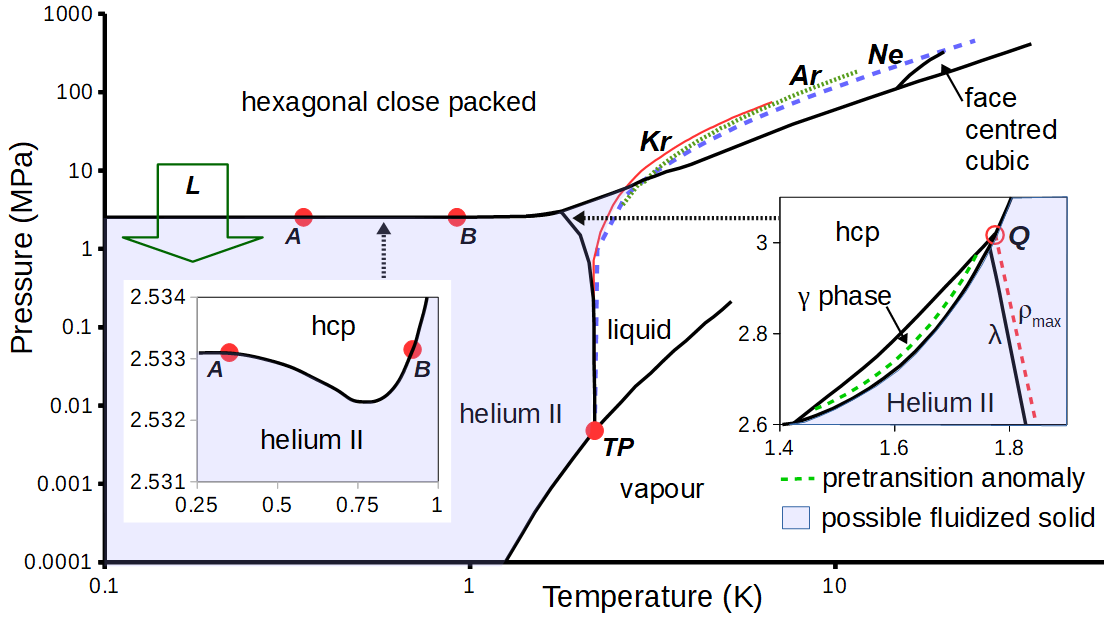}
	
	\caption{\em The phase diagram of \he4 on a logarithmic scale, with linear insets\cite{young1975phase,donnelly1998observed,brooks1977calculated,hoffer1976thermodynamic}.
		The melting curves of neon, argon and krypton are superposed
		with pressures and temperatures
		scaled so their triple points fall at `TP'.
		The $ \gamma $ phase (right inset)
		is currently classified bcc\cite{schuch1962structure}
		and exhibits a pretransition specific heat anomaly  (dotted)\cite{hoffer1976thermodynamic}.
	\singlequote{L} is the path discussed in the text.}
	\label{fig:phase-diagram-he4}
\end{figure}

London suggested how to resolve this anomaly in 1936 by showing
that cold helium II
has negligible entropy,
indicating a regular atomic arrangement.
He proposed a crystal-like lattice with enough free volume to allow atoms to move relative to one another,
which he associated with the low viscosity flow\cite{london1936condensed,frohlich1937lambda,london1939state}.
Compare Andreev and Lifshitz's 1969 proposal that 
atoms may advance through crystals,
making them ``neither a solid nor a liquid''\cite{andreev1969quantum},
and Leggett's 1970 proposal for supersolid flow\cite{leggett1970can}.

In general, solidification occurs when atoms or molecules cohere,
but helium atoms repel, assisted by 
zero point motion.
London argued that this accounts for the
transition near \singlequote{L} in figure \ref{fig:phase-diagram-he4}, which
``depends essentially'' on
volume rather than temperature\cite{london1939state}.
Advancing along \singlequote{L}, the distance between the atoms in solid helium increases
until they lose cohesion and fluidize.
Compare dry sand, which flows when the grains
separate and lose cohesion.

Experiments in another field provide quantitative 
evidence for this description.
Spherical colloidal particles in a fluid medium also repel, assisted by Brownian motion,
and form a hexagonal solid
which loses cohesion when the volume 
per particle is increased by diluting the suspension\cite{pusey1986phase,pusey2009hard}.
At intermediate dilutions
the colloidal solid is dispersed in 
a flowing phase which occupies a factor
of 1.103 more volume 
(the volume fractions being 0.545 and 0.494 respectively)\cite{pusey1986phase}.
This mirrors \he4
at constant volume and low temperature,
where a hexagonal close packed (hcp)
solid is dispersed in helium II
with the same volume ratio (figure \ref{fig:molar-volume})\cite{hoffer1976thermodynamic}.

\begin{figure}[htb]
	\centering
	
	\includegraphics[width=.75\linewidth]{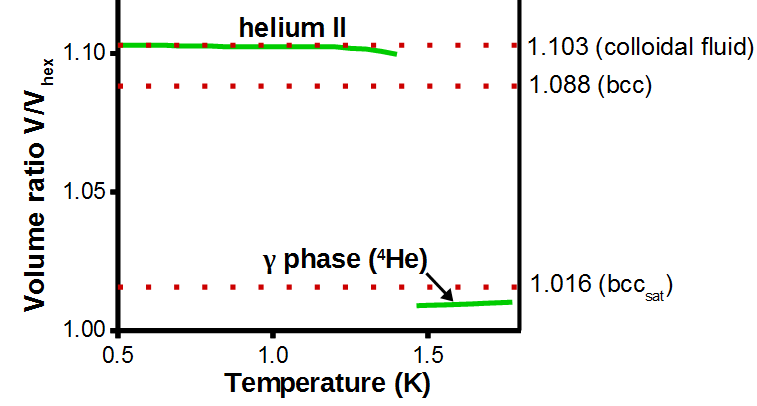}
	
	\caption{\em Experimental molar volumes $ V $
		(solid lines),
		plotted as $ V/V_{hex} $ where $ V_{hex} $ is the volume of the hexagonal solid
		under the same conditions\cite{hoffer1976thermodynamic,pusey2009hard}.
		1.103: flowing phase of spherical particles suspended in a fluid medium, 
		1.088: bcc
		lattice of hard spheres
		(packing efficiencies in the supplementary material), 
		1.016: bcc lattice
		saturated with interstitial atoms,
		discussed in the text.
	}
	\label{fig:molar-volume}
\end{figure}

London's 1936 description places the triple point of \he4 at \singlequote{TP} in figure \ref{fig:phase-diagram-he4}.
This quantitatively agrees with the other noble gases,
whose melting curves align in dimensionless units
of temperature and pressure relative to their
triple points.
Figure \ref{fig:phase-diagram-he4} superposes
the melting curves of neon, argon and krypton after scaling the temperatures and pressures
so their triple points fall at \singlequote{TP}.
They align with each other and with \he4 at higher temperatures (the small differences depend on
the square root of the atomic weight; see the supplementary material).
At lower temperatures, the solid phases of the other noble gases are face centred cubic,
a maximally efficient packing arrangement
with insufficient volume to fluidize;
this may be favoured by the occupied 
$ p $, $ d $ or $ f $ electron orbitals, vacant in \he4.
The scaled melting curves in figure \ref{fig:phase-diagram-he4} indicate approximately 
where London's fluidized solid (shaded) melts into an ordinary liquid. 

London suggested a diamond lattice as one
possibility\cite{london1936condensed}.
Fr\"ohlich then showed this is equivalent to a body centred cubic (bcc) lattice with 50\% vacancies, giving ample free volume to support the flow\cite{frohlich1937lambda}.
However, in 1938 London showed a diamond lattice is not viable because the vacancies are not stable\cite{london1938lambda,london1939state}.
He sidestepped the unsolved problem of
the precise geometry in order
to ``direct attention'' to a momentum space representation
in which the flow is quantized\cite{london1938lambda}.
We will show that his two descriptions,
in physical space (1936)\cite{london1936condensed} and momentum space (1938)\cite{london1938lambda},
are complementary.
More specifically, helium atoms
advance through interstitial positions in a locally bcc lattice, and interact via lattice vibrations
to produce coherent and quantized flow.

\mySection{$ \gamma $ phase of \he4} \label{sec:gamma-phase}

We begin with $ \gamma $ helium,
a crystalline solid (figure \ref{fig:phase-diagram-he4} inset).
Figure \ref{fig:molar-volume} shows that it
occupies approximately 7\% less molar volume than expected
for bcc crystals in a hard sphere approximation.
Nevertheless Schuch and Mills claimed in 1962
that this ``provides evidence that the $ \gamma $ phase is bcc''\cite{schuch1962structure}.
They noted that the hcp-$ \gamma $ volume ratio
in figure \ref{fig:molar-volume}
is approximately the same
as for the hcp-$ \beta $ transition in zirconium\cite{schuch1961bond,zhang2005experimental}, presumed that $ \beta $ zirconium is bcc, 
and cited Pauling's speculation
that metals have bonds
with covalent character whose lengths depend on the crystal geometry (coordination number)\cite{pauling1947atomic}.
But the original paper on the coincidence 
admitted that the bonding is not the same in metals and helium\cite{schuch1961bond},
and therefore Pauling's speculation about 
covalent bond lengths does not fully explain   
the anomalous molar volume of $ \gamma $ helium.

Schuch and Mills also reported x-ray measurements
which likewise exhibit anomalies that
are not explained
by classifying $ \gamma $ helium as ordinary bcc crystals.
The hcp phase has eight visible reflections while single crystals of the presumed bcc phases of both \he3 and \he4 have only three,
whose intensities decline steeply with increasing angle,
and the pattern for a coarse powder exhibits only one ill-defined reflection\cite{schuch1958structure,schuch1962structure}.
Schuch {\em et al} indexed the reflection angles 
to those of bcc crystals and dismissed the anomalous intensities as due to
zero point motion.
However, this does not account for the difference 
between a coarse powder and single crystals,
or hcp helium having more reflections
even though the reflection planes are closer.
Based on later measurements, 
zero point motion 
would attenuate the (200) line  by
approximately 50\% relative to the (110) line,
much less than observed 
(calculated in the supplementary material from the steep potential energy barrier 
to atomic displacements at approximately $ 0.2s $ where $ s $ is the length of a bcc cell\cite{moroshkin2008atomic}).

We interpret the anomalous molar volume and x-ray intensities as evidence that
the bcc crystals of $ \gamma $ helium
are saturated with interstitial atoms. 
Figure \ref{fig:bcc-to-fcc}a shows an effectively infinite bcc lattice of helium atoms, 
with one extra atom added, that has 
been relaxed to equilibrium
at low temperature in numerical simulation.
The resulting interstitial defect occupies  seven bcc cells, 
with the eight numbered atoms displaced as in \ref{fig:bcc-to-fcc}b.
Figure \ref{fig:bcc-to-fcc}c is  
a superlattice of defects,
or bcc lattice saturated with interstitial atoms.

\begin{figure}[htb]
	\centering
	\includegraphics[width=.55\linewidth]{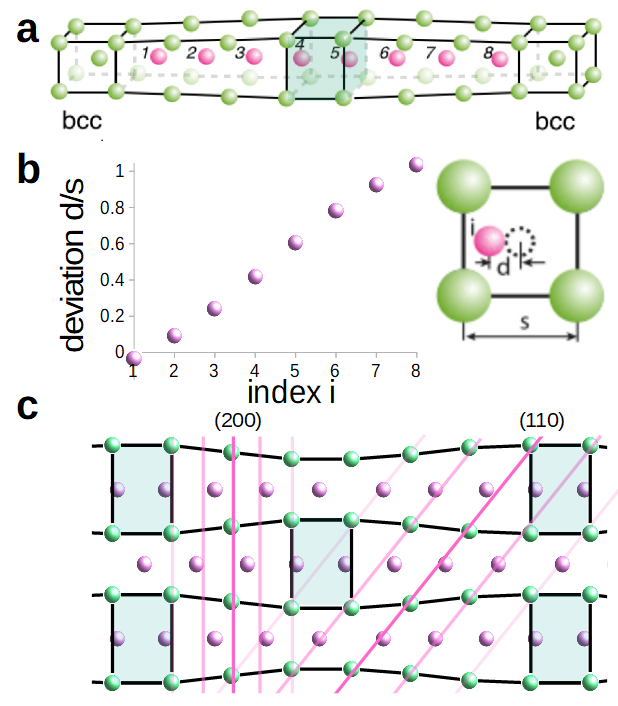}
\caption{\em 
		(a) Numerical simulation of an extra (interstitial) atom in a bcc lattice of \he4 atoms at 1 atmosphere and low temperature.
		Only significantly perturbed regions are shown.
		Program detail in supplementary material.
		(b) Displacement of atoms in the central row
(c) Superlattice of interstitial defects.
}
	\label{fig:bcc-to-fcc}
\end{figure}

This proposed structure for $ \gamma $ helium agrees with experiment in the following ways.

\begin{enumerate}[{A}1]

\item Interstitial atoms provide a ready supply of helium atoms at grain boundaries, 
which helps develop large crystals. 
Most samples formed single crystals, and 
a fine powder could not be made\cite{schuch1958structure,schuch1962structure}. 

\item An unperturbed bcc cell contains two atoms,
so the idealised arrangement in figure \ref{fig:bcc-to-fcc}c 
will occupy approximately $ \frac{14}{15} $ the molar volume of bcc crystals. 
The resulting hcp-$ \gamma $ volume ratio
is near the experimental value
(figure \ref{fig:molar-volume})\cite{hoffer1976thermodynamic}.

\item  \label{A:entropy-solid} Thermal agitation will increase the 
longitudinal separation between interstitial atoms in figure \ref{fig:bcc-to-fcc}c, so the hcp-$ \gamma $ volume ratio will increase with temperature, as observed\cite{hoffer1976thermodynamic}. The associated randomness raises the entropy
of the $ \gamma $ phase,
which will form on heating,
not cooling, hcp helium, also as observed.

\item The atoms in figure \ref{fig:bcc-to-fcc}c are significantly perturbed from a bcc arrangement,
so the intensities of their x-ray reflections will 
decline steeply with increasing angle,
where the reflection planes
are closer, giving fewer visible reflections than for hcp crystals.
The horizontal lines of atoms will produce weakened (200) reflections from single crystals.
However, for the vertical (200) reflection planes (marked), 
the added atoms alter the periodicity of alternate rows. This will mask the (200) reflections in a coarse powder,
where the orientations are random.
All these features were observed\cite{schuch1958structure,schuch1962structure}.

\item \label{A:gibbs-free-energy} 
Rearranging an atom into an interstitial position
changes the Gibbs free energy by
\begin{equation}\label{eq:gibbs-free-energy}
\Delta G = \Delta U + P \Delta V - T \Delta S
\end{equation}
where $ P $ is the pressure, $ T $ temperature,
$ \Delta U $ the change in internal energy,
$ \Delta V $ volume and
and $ \Delta S $ entropy.
This becomes negative at high enough pressure,
since $ \Delta V $ is negative.
Thus $ \gamma $ helium forms on increasing the pressure,
as observed (figure \ref{fig:phase-diagram-he4}).

\item \label{A:pretransition} Reducing the pressure of $ \gamma $ helium until 
$ \Delta G $ in \eqref{eq:gibbs-free-energy} vanishes will expel some interstitial defects,
but those that remain will be more dilute, with more entropy, which will stabilise them.
Thus the transition will not be sharp. 
This is observed as a substantial rise in the specific heat capacity
within 20mK of the melting temperature,
a previously unexplained `pretransition anomaly' (figure \ref{fig:phase-diagram-he4} inset)\cite{hoffer1976thermodynamic}.


\end{enumerate}

See the supplementary material for further evidence.

\mySection{Rotons}

When the Gibbs free energy \eqref{eq:gibbs-free-energy} changes sign on reducing the pressure,
interstitial atoms will be expelled from $ \gamma $ helium.
This suggests the new phase is locally bcc.
Figure \ref{fig:phase-diagram-he4} identifies it
as helium II, and figure \ref{fig:molar-volume} shows it has approximately the expected molar volume.
Interstitial atoms may advance
through it as shown in figure \ref{fig:moving-defect} (animation in supplementary material). 
These mobile interstitial atoms
have the properties attributed to the excitations known as rotons\cite{feynman1954heliumtwofluid}, as follows.

\begin{figure}[htb]
	\centering
	\includegraphics[width=.45\linewidth]{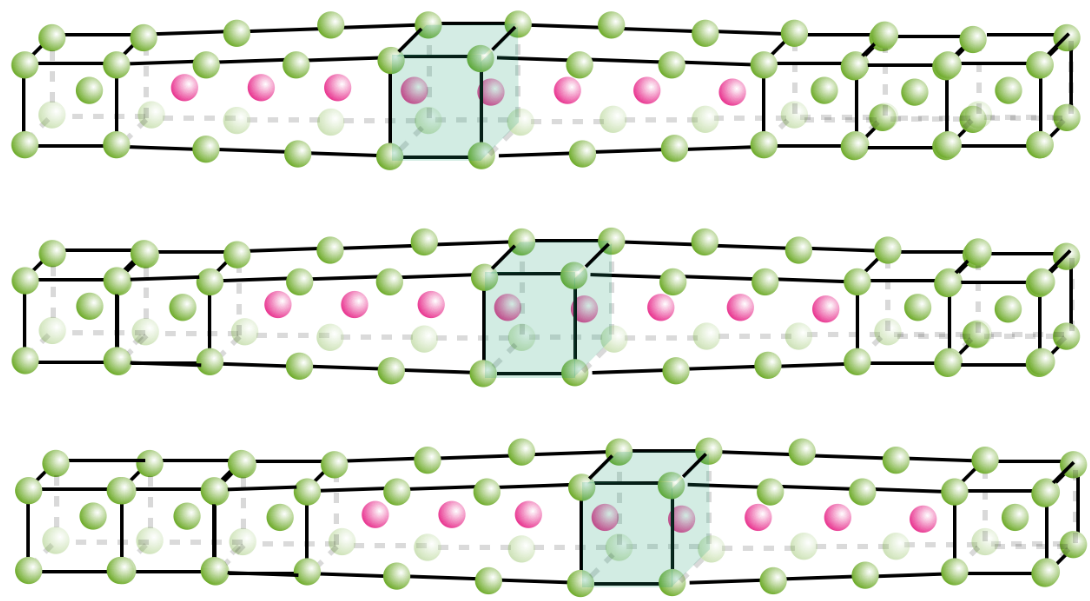}	
	\caption{\em 
		An interstitial atom advancing through an
		idealised bcc lattice.
		Numerical calculation at low temperature and 1 atmosphere pressure. 
	}
	\label{fig:moving-defect}
\end{figure}

\begin{enumerate}[{B}1]

\item \label{B:tucker-wyatt-experiment}
In 1999, Tucker and Wyatt
created excitations in cold helium II 
using a pulse of heat, which advanced 
linearly until reaching the surface 
and ejecting a helium atom 
into the space above\cite{tucker99antiparallel}.
By directing them at an angle and observing
the ejected atoms, they showed that the excitations have positive effective mass.
These observations are consistent with interstitial atoms as in figure \ref{fig:moving-defect}. 

\item \label{B:tucker-wyatt}
Tucker and Wyatt argued that excitations with nonzero effective mass 
do not lose energy to acoustic waves, 
as the energy and momentum changes 
cannot match simultaneously.
Interstitial atoms have the same property.
They observed linear motion without noticeable losses\cite{tucker99antiparallel}.

\item \label{B:vacancy} Vacancy defects resemble figure \ref{fig:moving-defect} with a missing atom
instead of an added one, and move in a similar way.
They have negative effective mass,
and more enthalpy since they increase the volume against the pressure.
By colliding excitations together, Tucker and Wyatt discovered mobile excitations 
which they showed have negative effective mass\cite{tucker99antiparallel}. They were not created by the pulse of heat,
suggesting they have greater enthalpy.

\item In the 1940's, 
Landau concluded from phonon dispersion data that helium II 
contains excitations\cite{landau1947theory}.
There are two species, 
$ R^+ $ (rotons) and $ R^- $ (maxons),
with positive and negative effective mass respectively.
Interstitial atoms and vacancies have these properties.

\item \label{B:conserved-atoms} When negative ions are drawn through helium II
under an electric field and  
scatter inelastically, they will 
create interstitial atoms and vacancies together, since atoms are conserved. 
In 1976, Allum, Bowley and McClintock 
discovered ``hitherto unrecognized selection rules whereby
rotons are only created in pairs''
in this experiment\cite{allum1976rotonpair}.

\item \label{B:effective-mass} The conventional model of rotons does not predict their effective mass.
If the $ i $'th numbered atom in figure \ref{fig:bcc-to-fcc}a
advances at velocity $ v_i $, 
the kinetic energy will be $ T \approx \frac12 m \Sigma v_i^2$
where $ m $ is the atomic mass. 
Our numerical model at atmospheric pressure
and low temperature indicates
$ T = \frac12 m^* v^2$ where $ v $ the velocity of the defect
and $m^* \approx 0.165m $ (see the supplementary material).
This is within 5\% of the observed value\cite{brooks1977calculated}.

\item In 1954, Feynman suggested that 
a roton's kinetic energy is due to rotating flow patterns\cite{feynman1954heliumtwofluid}.
This suggests that its effective mass
is greater than its gravimetric mass (if any),
unlike an interstitial atom which has the mass of a helium atom.
This difference can be studied using density data.
When dilute,
interstitial defects move freely in one dimension, and their concentration can be
calculated from their quantum wavelength
similarly to a particle in a box.
At temperature $ T $, the expectation number $ \langle N \rangle $ in a bcc lattice of $ N_o $ atoms is given by 
\begin{equation}\label{eq:density}
\frac{\langle N \rangle}{N_0} ~~=~~ 3
\, \left (\frac{2 \pi m^* s^2}{h^2 \beta}\right )^\frac12 
~ e^{-\beta \Delta H}
\end{equation}
where $ s $ is the length of a bcc cell, $ h $ is Planck's constant, 
$ \Delta H = \Delta U + P \Delta V $ is the enthalpy of a defect and
$ \beta = (K_B T)^{-1} $ where $ K_B $ is Boltzmann's constant.
See the supplementary material for the proof and discussion of small terms we have neglected.
These interstitial defects increase the density of helium II,
which is known from dielectric observations,
and figure \ref{fig:activation-temperature}
shows good agreement with experiment spanning
six orders of magnitude\cite{brooks1977calculated}.


\begin{figure}[htb]
	\centering
	\includegraphics[width=0.8\linewidth]{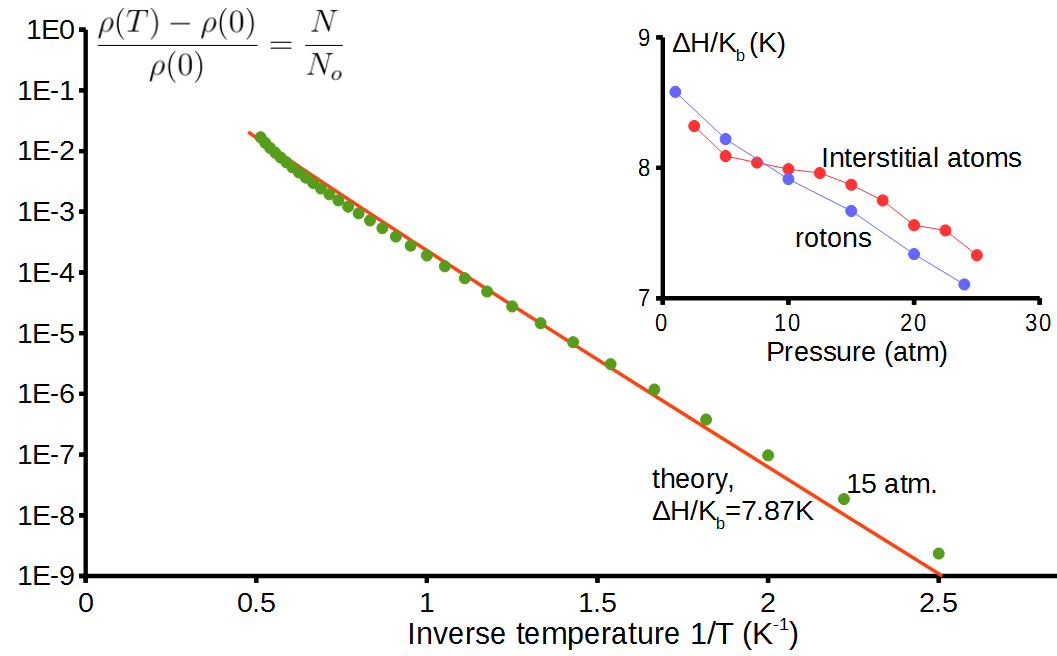}
	\caption{\em Concentration of interstitial atoms in helium II,
		from the measured density $ \rho(T) $
		at 15 atmospheres pressure
		after subtracting out ordinary (Debeye) expansion\cite{brooks1977calculated},
		compared to equation \eqref{eq:density}.
		Inset -- activation temperature as a function of pressure.
		The roton energy gap is from neutron scattering data at the lowest temperature measured, approximately 1.25K\cite{brooks1977calculated}.}
	\label{fig:activation-temperature}

\end{figure}

\item The enthalpy $ \Delta H $ of
interstitial atoms, obtained above,
is within 5\% of the
roton energy gap at all pressures (figure \ref{fig:activation-temperature} inset)\cite{brooks1977calculated}.

\item Hcp helium
has insufficient volume for interstitial atoms,
and vacancies have a large excitation energy (see B\ref{B:vacancy}).
This will suppress supersolid flow in this phase (unlike helium II, which has more room for interstitial atoms).
Attempts to demonstrate supersolidity in hcp helium were unsuccessful\cite{kim2004probable,maris2012effect,balibar2016dislocations}
but it may have been seen in
other systems with looser crystal-like structures\cite{li2016observation,leonard2016supersolid}.

\end{enumerate}

See the supplementary material for further evidence.

\mySection{Fluidization mechanisms in `solid' helium II}

External stresses will locally raise the pressure
at protrusions in the surface of the bcc crystals of helium II.
This reverses the sign of $ \Delta G $ in \eqref{eq:gibbs-free-energy}
so that interstitial atoms form, which 
advance through the solid and 
contribute to the flow.
The flow will be amplified because interstitial atoms 
are conserved (see B\ref{B:conserved-atoms})
and move without resistance between collisions (B\ref{B:tucker-wyatt}).
There may also be other flow mechanisms,
resembling the giant plasticity of hcp helium\cite{haziot2013giant}.

This agrees with experiment as follows.

\begin{enumerate}[{C}1]

%

	\item \label{C:bcc-correlation-length} 
	The density of helium II at its  
	saturated vapour pressure and 1.1K is 27.84g/mole\cite{brooks1977calculated}.
	Bcc crystals with this density
	would have a (110) peak in their structure factor at 19.7nm\textsuperscript{-1}.
	Neutron scattering measurements\cite{caupin2008static}
	indicate the peak is at 20.3nm\textsuperscript{-1}.


\item \label{C:damage-to-lattice} 
We saw (figure \ref{fig:molar-volume}) that colloids have phases resembling helium.
 The crystalline phases of colloids are delicate; for example, they are perturbed by gravity and substantially damaged under shear\cite{pusey1986phase,pusey2009hard,zhu1997crystallization},
 and we would expect the crystal structure in helium II to be similarly damaged,
 for example by vibrations or flow.
Its structure factor has a half-width of approximately 15\%\cite{caupin2008static},
	indicating that the lattice maintains coherence 
	over distances of order the length of the interstitial defect in figure \ref{fig:bcc-to-fcc}a, but not much longer.

	\item \label{C:volume} The perturbations described above will 
	impair the mean packing efficiency.
	Figure \ref{fig:molar-volume} shows that 
	helium II occupies 1.4\% more volume than for
	perfect bcc crystals on a hard sphere approximation\cite{hoffer1976thermodynamic}.

	\item \label{C:solidify-heating-he4}
	The factors driving the pretransition anomaly  in
	$ \gamma $ helium
	(see A\ref{A:pretransition}) also apply elsewhere
	on the melting curve.
	In particular, near \singlequote{A} in figure \ref{fig:phase-diagram-he4},
	helium II has few interstitial atoms, from \eqref{eq:density}, and a $ \gamma $-like 
	pretransition phase would have many more, giving it more entropy when warmed (see A\ref{A:entropy-solid}).
	Thus the solid will form on heating, as observed.
	On further heating,
	the exponential rise in \eqref{eq:density}
	will reverse the entropy balance and the solid will re-melt,
	observed at \singlequote{B}.
	
\end{enumerate}

See the supplementary material for further evidence, and comparison with current 
models.

\mySection{Lattice vibrations} 

The equation of motion
for a uniform line of atoms in one dimension
has wave-like solutions (phonons) up to a frequency $ f_o $ where neighbouring atoms oscillate antiphase\cite{donovan1971lattice}.
The supplementary material shows that 
$ f_o \approx 160$GHz in cold helium II at atmospheric pressure.

Near a discontinuity such as an interstitial defect,
there is a so-called `optical' solution  
just above $ f_o $
where the unnormalised displacement
of the $ n $'th atom at position $ x_n $ 
in the one-dimensional lattice
is
\begin{equation}\label{eq:moving-resonance}
	U_n ~~ \approx ~~
	(-1)^n
	~ e^{\nu t -\mu x_n}
	~ \cos(k x_n - \omega t)
\end{equation}
See the supplementary material for the proof and 
extension to three dimensions
involving spherical harmonics.
The amplitude $ e^{\nu t - \mu x_n} $
decays with distance from the defect,
where both  parameters reverse sign.
By inspection, it advances at velocity $ v = \nu/\mu $, which we associate with the velocity of the defect.
The wavevector is 
$ k = -\omega v  / c_s^2 $ where $ c_s $ is the speed of sound.

This resonance will be excited by the energetic
processes that create an interstitial atom,
and it is long-lived since propagating waves
do not exist at the resonant frequency
and cannot carry energy away.
Thus, a newly created interstitial atom is an association between a particle and a wave.
The wave will guide the trajectory of the interstitial atom, 
since any velocity changes
would require changes to the wave modes.
This is consistent with the trajectories  
observed by Tucker and Wyatt (B\ref{B:tucker-wyatt})\cite{tucker99antiparallel},
which were ballistic on distances significantly longer than the 
coherence length of the lattice.

A similar phenomenon occurs in another association between
a particle and a wave, 
a droplet of oil bouncing on a vibrating oil tray. 
The bouncing creates surface waves which guide or `pilot'
the droplet as it moves across the surface
in a so-called `path memory' effect,
which produces ballistic trajectories 
as with interstitial atoms\cite{fort2010path}.
If barriers are present, such as a pair of slits,
the trajectories exhibit statistical diffraction and interference patterns 
resembling those of a third wave-particle association, a quantum particle\cite{couder2006single,eddi2009unpredictable,fort2010path,bush2015new,bradyanderson2014why}.

If there are multiple interstitial atoms, their resonant modes will overlap and weakly couple.
Coupled resonators have normal modes with raised and lowered frequencies,
and they spontaneously synchronize when 
one of them is selected for,
a phenomenon of nonlinear origin
first noticed in pendulum clocks in 1665
and now studied in the field of Kuramoto theory\cite{acebron2005kuramoto,bennett2002huygens}.
The alignment can be maximised when the resonators are separated by a fixed number of wavelengths;
this produces coherent motion  
which is described by a shared order parameter $ \Delta(\textbf{x}, t) $.
Such coherence is also seen in videos of oil drops,
which move coherently across the surface, 
separated by a fixed number of wavelengths\cite{filoux2015strings}.

When interstitial atoms are synchronized in this way,
they will not collide with each other,
or even their images in the boundary.
We saw (B\ref{B:tucker-wyatt}) that they move between collisions without noticeable loss\cite{tucker99antiparallel},
and so the flow they carry will also be without noticeable loss.
We associate this with the superflow in helium II\cite{leggett2006quantum}.

There is a fourth particle-wave association,
which exhibits a similar order parameter.
In 1957, Bardeen, Cooper and Schrieffer (BCS)\cite{bardeen1957theory}
showed that conduction electrons
in a metal are associated with acoustic waves in the lattice,
which they represented as virtual phonons.
If the wave frequency is $ f $,
they showed that the electrons synchronize  
when their energy levels differ 
by less than $ h f $.
Josephson showed how to measure the phase
of the associated order parameter  $ \Delta(\textbf{x}, t) $ in 1962\cite{josephson1962coupled}.
When in this coherent state, the electrons also move without resistance.

In BCS theory, the wavelength of the order parameter for the electrons is $ \lambda=h/p $
where $ p $ is the momentum of a pair of electrons\cite{josephson1962coupled,leggett2006quantum}.
Compare a pair of interstitial helium atoms,
whose momentum is $ 2 m^* v $ .
The solution \eqref{eq:moving-resonance}, and its associated order parameter, have wavelength
\begin{equation}\label{eq:wavelength}
	\lambda 
	~~=~~ \frac{2 \pi}{|k|}
	~~=~~ \frac{c_s^2}{f v}
	~~=~~ \frac{h_s}{2 m^*v}
\end{equation}
where we have defined $ f=\omega / 2 \pi $ and 
$ h_s = 2 m^* c_s^2/f $.
Figure \ref{fig:planck-constant} shows that 
$ h_s $ is the same as Planck's constant within experimental accuracy
at all pressures\cite{brooks1977calculated}.
The reason for this empirical agreement 
with the BCS theory of superconductivity
remains to be understood.

\begin{figure}[htb]
	\centering
	\includegraphics[width=.5\linewidth]{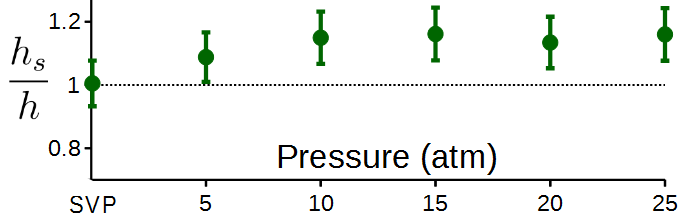}
	
	\caption{
		\em The parameter $ h_s = 2 m^* c_s^2/f $ in \eqref{eq:wavelength}
		divided by Planck's constant,
		from data for helium II at low temperature\cite{brooks1977calculated}.
		Error bars correspond to different crystal directions and do not include 
		experimental error,
		the damage to the bcc lattice due to the flow, or our approximation $ f \approx f_o $ (see the supplementary material).
	}
	\label{fig:planck-constant}
\end{figure}

%

\he3 differs from \he4 in having magnetised nuclei.
By inspection of figure \ref{fig:moving-defect}
or the animation in the supplementary material,
each atom in the path of a moving defect 
advances by one
bcc cell's length.
This will disturb the correlations among the magnetised nuclei, giving dissipative flow
as observed.
This source of dissipation will be quenched at millikelvin
temperatures, where the nuclear spins 
in a bcc lattice become aligned\cite{panczyk1967evidence,hoffer1976thermodynamic}.
\he3 is observed to become superfluid at these temperatures.

\mySection{Mesophases}

Rod-shaped colloidal particles dispersed in a fluid medium
form mesophases (liquid crystals),
which are intermediate between a liquid and a solid.
When dilute, the rods are randomly oriented (isotropic),
and at greater concentration they align parallel to one another (nematic),
which minimizes repulsive inter-particle excluded volume
interactions\cite{onsager1949effects,degennes1974liquidcrystals}.
At intermediate concentrations there 
is a biphase of isotropic and nematic domains\cite{lekkerkerker2013liquid}.

Similar considerations apply to the
rod-shaped interstitial defects in helium II,
which are locally aligned with the
bcc lattice.
At low temperature they are dilute and
populate all three local directions of the bcc lattice randomly (isotropically).
When warmed, the concentration rises
(figure \ref{fig:activation-temperature}),
which forces more of them into nematic domains,
where they are closely correlated, 
giving additional modes to shed energy and momentum,
which introduces dissipation similar to a normal liquid.
This agrees with the empirical
two-fluid model of helium II, 
where it is superfluid at low temperature
with an increasing proportion of normal fluid 
when warmed\cite{balibar2014superfluidity}.

At \singlequote{$ \lambda $} in figure \ref{fig:phase-diagram-he4},
the isotropic domains become isolated, 
quenching the superflow on long distance scales.
Just above this temperature, the 
isotropic domains disappear.
It is more difficult to increase the concentration of repulsive rod-like defects in a nematic phase,
where the packing is more efficient,
accounting for the sudden fall in the specific heat capacity and
the reversal in the thermal expansion coefficient
at \singlequote{$ \rho_{max} $} in figure \ref{fig:phase-diagram-he4} inset.
This line ends at a quadruple point (\singlequote{Q}) where four boundaries meet\cite{brooks1977calculated}. 
Each boundary imposes a constraint on the variables pressure and temperature,
which would exceed the available degrees of freedom if there were one component
(the Gibbs phase rule).
\singlequote{Q} can exist because there are two components, the lattice and
interstitial defects.

\mySection{Further research}

Helium is not the only system where an unusual flowing phase
coexists under pressure with abnormally 
large crystals that are presumed to be bcc. 
The Earth's core and neutron stars are thought to
have these characteristics\cite{kobyakov2014towards,vocadlo2003possible},
raising the possibility that the crystal structures and flow mechanisms are related.
Helium II has a non-stochiometric bcc-like structure, due to the interstitial atoms,
as do high temperature superconductors such as sulfur hydride at high pressure\cite{errea2016quantum},
raising the possibility that their flow mechanisms
are similar.

\ifsupplementary


%
%
%

\newcommand{\refFigMolarVolume}{2}
\newcommand{\refFigBccToFcc}{3}
\newcommand{\refFigActivationTemperature}{5}
\newcommand{\refFigPlanckConstant}{6}
\newcommand{\BeffectiveMass}{B6}
\newcommand{\figPhaseDiagramHeFour}{1}
\newcommand{\AgibbsFreeEnergy}{A5}
\newcommand{\AentropySolid}{A3}
\newcommand{\Apretransition}{A6}
\newcommand{\eqDensity}{(2, text)}
\newcommand{\figActivationTemperature}{5}
\newcommand{\figPlanckConstant}{6}
\newcommand{\CdamageToLattice}{C2}
	\newpage
	\mySection{\large Supplementary Material}
	\label{sec:supplementary-material}

%
%

	\mySection{File list}
	
	The following files are found at 
	\\ {https://drive.google.com/drive/folders/0B-zmIlkqbDkZX0owZXJyajc4alE?usp=sharing}.
	
	\newcommand{\figMolarVolume}{2}
	
	\begin{tabular}{lll}
		File  & Description
		\\\hline
		interstitial-atom-movie.gif & Animation of moving interstitial atom			
		\\ densityhelium.ods$ \dagger $  & Molar volumes and forces between helium atoms
		\\interstitial.ods$ \dagger $  & Renders and analyses the output of the program
		\\interstitial2.ods$ \dagger $ & 2-D rendering of interstitial atom
\\interstitial-movie.ods$ \dagger $ & spreadsheet for the animation
		\\moving-defect.odg*  & flow mechanism
		\\structure-factor.ods$ \dagger $ 
		& Width of the structure factor of helium II
		\\	thermal-expansion.ods$ \dagger $  & The thermal expansion coefficients
		\\	triple-points.ods$ \dagger $ & The phase diagrams of the  noble gases\\\hline
	\end{tabular}

$ \dagger $ LibreOffice version 5.2 spreadsheet.
* LibreOffice version 5.2 drawing

\newcommand{\figBccToFcc}{3}
The program to calculate the equilibrium positions of the atoms
near a defect in figure \figBccToFcc{} is in the `program' folder.
It uses Microsoft Visual Studio 2015.

\mySection{Melting curves of the noble gases}

Figure \figPhaseDiagramHeFour{} shows that the melting pressures and temperatures
of the noble gases, including \he4,
approximately coincide at warmer temperatures
when plotted in 
dimensionless units of temperature and pressure
relative to their triple point values.

The small differences depend systematically on the atomic weights.
Figure \ref{fig:sqrt-atomic-weight} 
shows the dimensionless melting pressures
of helium, neon and argon,
at a temperature of $ 4 T_{triple} $,
plotted against $ \sqrt{W} $
where $ W $ is the atomic weight.
Note the linear trend line, with which \he4 is in good agreement when its triple point is placed at \singlequote{TP} in figure
\figPhaseDiagramHeFour{}.
	
\begin{figure}[htb]
	\centering
	
	\includegraphics[width=.55\linewidth]{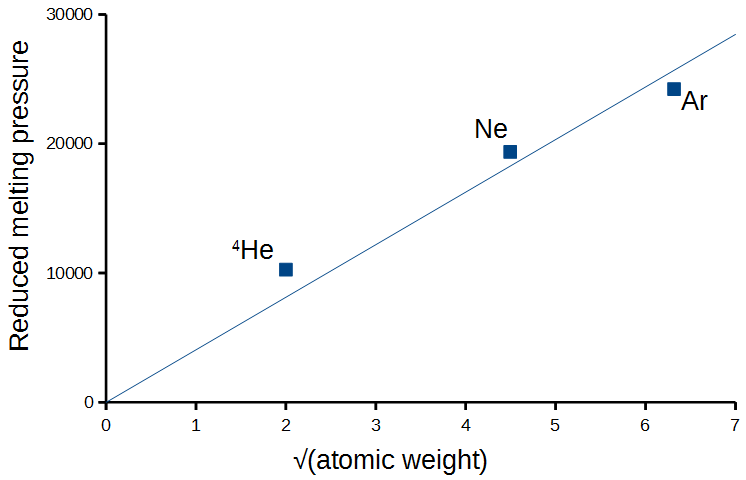}
	
	\caption{\em The reduced melting pressures
		$ P/P_{triple} $ of argon, neon and \he4
		at 4 times the triple point temperature.
		\he4 is in good agreement with the trend when its triple point is placed at \singlequote{TP} in figure \figPhaseDiagramHeFour{}\cite{young1975phase,donnelly1998observed,brooks1977calculated,hoffer1976thermodynamic}.
	}
	\label{fig:sqrt-atomic-weight}
\end{figure}

\mySection{Packing efficiencies}

The following table summarises
the packing efficiencies of the geometric arrangements discussed in the text,
using a hard sphere approximation
unless otherwise stated.

\noindent \begin{tabular}{|l|l|l|l|}
	\hline		
	Geometric arrangement					&	efficiency				& Our classification		& Literature
	\\\hline
		face centred cubic lattice										& 0.74 								& fcc solid	& fcc solid
	\\hcp lattice										& 0.74 								& hexagonal solid		& hexagonal solid
	\\ bcc saturated with interstitial atoms& 0.729$ ^\dagger $							& $ \gamma $ phase of \he4		& --					
	\\ bcc lattice					& 0.68								& --					& $ \gamma $ phase of \he4
	\\ fluidized bcc crystals 						& 0.67\textsuperscript{*}								& helium II				& --
	\\	quantum fluid 						& 0.67\textsuperscript{*}									& --					& helium II: London 1938
	\\ diamond lattice & 0.34 & &helium II: London 1936
	\\\hline
\end{tabular}

\noindent $ \dagger $Idealised arrangement in figure \figBccToFcc{}c ~ *Experimental value from helium II and 
spherical colloids.

\mySection{X-ray attenuation due to zero point motion}

We now estimate  the attenuation of the intensities of
the x-ray diffraction lines in hypothetically perfect
bcc crystal of helium atoms,
caused by zero point motion.
We will use recent estimates of the 
potential energy of a helium atom,
which is in a shallow well with a steep potential energy barrier to displacements greater than $ d_o \approx 0.2s $ where $ s $ is the side of a bcc cell\cite{moroshkin2008atomic,aziz1995ab}.

We begin with the approximation that  the probability density of a helium atom is
constant inside a sphere of radius $ d_o $.
The (200) reflection planes are $ d_1 = 0.5s $ apart, so the amplitude of the diffraction line will be attenuated by a factor 
$ \frac3{4 \pi} \int_{-1}^1 \pi (1-x^2)  \, \cos(2\pi\, [d_o/d_1] x) \, dx \approx 0.5 $,
and the attenuation for the (110) line would be 0.8.

The square of this ratio, approximately 40\%,
is an estimate for the relative attenuation
of the two lines.
This is an over-estimate since we have assumed a constant probability density,
whereas the ground state would have a density peak near the centre.
This is much smaller than required to  account for the 
observed attenuation.
For example, in 1958 Schuch, Grilly and Mills found that the x-ray line intensities
from single crystals of \he3 `declined steeply with increasing angle'
so that only three lines could be observed,
and in a coarse powder only a single (110) line was visible\cite{schuch1958structure}.
The corresponding phase of \he4 is similar\cite{schuch1962structure}.

	\mySection{Numerical calculation} 
	The net inter-atomic force as a function of the distance between atoms 
	was estimated from the density of \he4 as a function of pressure at 0.1K\cite{brooks1977calculated} (see figure \ref{fig:atomic-force}).
	This force was then used as an input to the
	computer program, 
	which iterates through the atoms, relaxing them in the direction of any net forces, until the forces are small.
	The calculation can thus be classified as a mean-field or semi-classical approximation.
	
	\begin{figure}[htb]
		\centering
		
		\includegraphics[width=.55\linewidth]{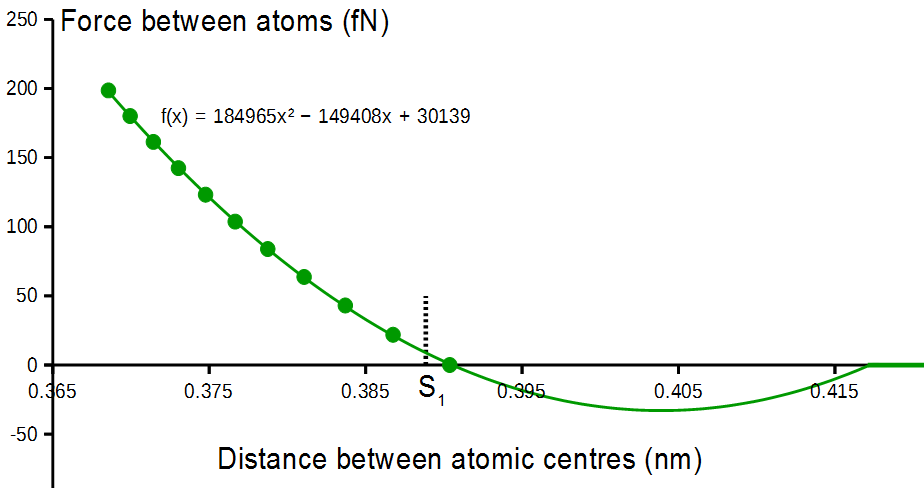}
		
		\caption{\em The net forces between neighbouring atoms, estimated from the density of \he4 as a function of pressure at 0.1K\cite{brooks1977calculated}.
			The line shows a quadratic fit to the data,
			which was used as an input to the calculation; the
			forces at larger distances were extrapolated as shown.
			$ S_1 $ is the distance between nearest neighbour atoms in an unperturbed bcc crystal at 1 atmosphere pressure
			used in figure \figBccToFcc{}.
			}
		\label{fig:atomic-force}
	\end{figure}

	In the starting conditions for the program, a bcc arrangement
	was set up in accordance with the observed density at the relevant pressure.
	An extra atom was added near the origin
	and the nearby atoms were displaced slightly 
	to assist convergence.
	The atoms were represented by objects in a collection, indexed by their unperturbed positions,
	so it was only necessary to store the objects that had been perturbed,
	thereby allowing computation with an effectively infinite 
	lattice.
	
	The file `positions.csv' in the program folder 
	contains the output of the program, including a large number of atoms which are
	perturbed very little.
	It has columns for the nominal positions (x0, y0, z0) 
	and relaxed positions (x, y, z)	of the atoms, in 
	descending order of the displacement from their nominal positions. 
	The unperturbed bcc lattice is 
	comprised of two intersecting cubic lattices, each of side 2 units;
	see the `coordinates' structure for detail.
	
	The spreadsheet `interstitial.ods' contains the coordinates for the defect plotted in figure \figBccToFcc{}.

	In our semi-classical or mean field approximation at low temperature, helium has the property, known since at least 1936,
	that the atoms can reduce their potential energy by becoming displaced
	slightly from the central position of a lattice\cite{london1936condensed}.
	The resulting Peierls-like distortion can be seen by careful examination of the program output
	(see for example the small asymmetry in figure \ref{fig:centre-row-displacements}),
	but it is small at 1 atmosphere pressure.

\mySection{$ \gamma $ helium: further experimental evidence}

Further experimental evidence that
$ \gamma $ helium has the structure shown
in figure \figBccToFcc{} is as follows.

\begin{enumerate}[{a}1]
	
%
	
	\item We saw (\AgibbsFreeEnergy) that interstitial atoms are forced into
	the bcc crystals in $ \gamma $ helium under pressure.
	At the transition pressure, we would expect the 
	energy $ P \Delta V $ to be comparable with
	other rearrangement energies involved. 
	Experimentally, $ P \Delta V \sim -10^{-22} $J,
	or 7K expressed as a temperature, comparable to
	the temperature required to melt the crystals in \he3 and \he4.
	
	\item Schuch and Mills reconciled the molar volume of $ \gamma $ helium with their x-ray measurements\cite{schuch1962structure}.
	However, they did not report the width of the reflection spots or the likely error in their estimate, even though they noted that the data were of poor quality. 
	When can estimate the scatter from their 
	measurements of the $ d110 $ distance,
	which corresponded to the brightest reflection.
	They averaged 11 photographs
	and reported an estimated error of 0.31\%.
	Assuming they used standard statistics, we can infer a scatter corresponding to 1\% 
	in their estimates of the positions of the centres of the individual spots,
	which translates into a scatter of 3\% in molar volume
	if one took the centres of the spots (which is itself an average).
	This implies that the width of the spots was substantially greater, consistent with the
	proposed arrangement in figure \figBccToFcc{}c.

	\item Schuch noted in 1961 that the 
	volume increase on the hcp-$ \beta $ transition in
	zirconium is proportionally the same as 
	for the hcp-$ \gamma $ transition in helium\cite{schuch1961bond}.
	Schuch and Mills later inferred that they must have the same structure\cite{schuch1962structure}.
	If so, $ \beta $ zirconium,
	like $ \gamma $ helium, is composed of bcc crystals saturated with intersitial atoms.
	This is supported by several observations.
	The transition occurs on heating, not cooling\cite{young1975phase},
	as expected from the entropy in the interstitial zirconium atoms (compare \AentropySolid{}).
	Like $ \gamma $ helium,
	it only exists over a narrow range of temperatures,
	it has an indistinct phase boundary on heating (compare \Apretransition{}), and 
	near the transition the intensities of its x-ray reflections decline steeply with
	increasing angle\cite{zhang2005experimental}.
	Heating produces a phase without sharp x-ray reflection peaks, resembling helium II
	(we would not expect it to fluidize
	as zirconium is not inert).
%
%

\end{enumerate}

\mySection{Effective mass} 
When an interstitial defect advances the length of a bcc cell, $ s $,
then the $ i $'th atom advances by $ (d_{i+1} - d_i) $ where $ d_i $ is the deviation shown in figure \figBccToFcc{}b.
Thus 
\[
	\frac{v_i}{v} ~~\approx ~~ \frac{d_{i+1} -d_i }{s}
\] where $ v_i $ is the velocity of the $ i $'th atom and $ v $ the
velocity of the defect. This is plotted in figure \ref{fig:centre-row-displacements}.

\begin{figure}[htb]
	\centering
	\includegraphics[width=.5\linewidth]{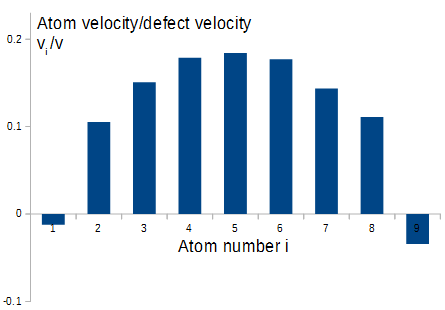}
	\caption{\em The velocity $ v_i $ of the $ i $'th atom of an advancing defect, divided by the velocity $ v $ of the defect.
		We attribute the small asymmetry
		and negative end values to the Peierls-like distortion discussed above. 
		The effective mass of the defect is $ m \sum (v_i/v)^2 $ where $ m $ is the mass of an atom.
	}
	\label{fig:centre-row-displacements}
\end{figure}

The kinetic energy of the atoms in the central row of the defect 
is $ T = \Sigma \frac12 m v_i^2 $.
Equating this to the kinetic energy of the defect moving at velocity $ v $ with effective mass $ m^* $, namely $ T = \frac12 m^* v^2 $, 
gives $ m^* = m \, \Sigma (v_i/v)^2 $. 
This sum is tabulated below, giving $ m^* = 0.165m $.
The sum $ \sum v_i/v $ differs from 1, indicating the error bar
due to the boundary of our calculation.

\begin{center}
	\spacing{1}
	\begin{tabular}[b]{lll}
		$i$ & $ v_i / v $ & $  (v_i/v)^2 $ 
		\\\hline
		1 & -0.0121 &	0.0001 \\
		2 & 0.1054 & 0.0111 \\
		3 & 0.1508	& 0.0227 \\
		4 & 0.1788	&0.0319 \\
		5 & 0.1841	&0.0339 \\
		6 & 0.177	&0.0313 \\
		7 & 0.1435	&0.0206 \\ 
		8 & 0.1109	&0.0123 \\
		9 & -0.0341	&0.0012 \\
		\hline 
		& 1.0042	& 0.1652 \\
		\hline
	\end{tabular}
	\spacing{2}
\end{center}

\mySection{Expectation number of defects} 

The partition function of a system with energy levels $ E_i $ in equilibrium 
at temperature $ T $ is defined by
\[
	Z~~=~~\sum e^{-\beta E_i}
\] 
where $ \beta = (K_B T)^{-1} $ and $ K_B $ is Boltzmann's constant.

Consider a single interstitial defect in one dimension,
which has effective mass $ m^* $
and momentum $ p $.
Its partition function is
like that of a particle 
in a one-dimensional box of length $ L $, namely
\[
	Z_{1}^{(1d)} ~~=~~\frac1h \int \int dq ~dp~ 
	e^{- \beta \frac{p^2}{2 m^*}}
	~~=~~\frac{L}{h} \, \sqrt{\frac{2 \pi m^*}{\beta}}
\]

If the centre row of the defect forms part of 
a line which (without the defect) contains $ M $ atoms,
separated by distance $ s $,
then substituting $ L = s M  $ gives 
\[
	Z_1^{(1d)} ~~ = ~~ M \, \sqrt{\frac{2 \pi m^* s^2}{\beta h^2}}
\]

Consider a large bcc crystal
containing $ N_o = k M $ atoms
in $ k $ identical lines of atoms like the above.
Neglecting edge effects, the defect
could occupy any one of the lines
and the partition function for defects oriented
parallel to the $ x $ direction will be 
\[
	Z_{1x}^{(3d)}
	~~=~~k Z_1^{(1d)}
	~~=~~ N_o \, \sqrt{\frac{2 \pi m^* s^2}{\beta h^2}}
\]

The defect could be oriented in the $ x $, $ y $ or $ z $ directions, giving the total partition function
\begin{equation}\label{eq:3d-partion-function}
		Z_1^{(3d)}~~=~~3 N_o \, \sqrt{\frac{2 \pi m^* s^2}{\beta h^2}}
\end{equation}
where $ s $ is the length of a bcc cell (the lattice parameter).

If the crystal contains $ N_i $ interstitial defects of chemical potential $ \mu $, the grand partition function is defined by
\[
	\mathcal{Z} ~~=~~\sum e^{\beta (N_i \mu - E_i)}
\]
which evaluates to
\[
	\mathcal{Z} ~~=~~ 1
	~+~ Z_1 \, e^{\beta \mu} 
	~+~ \frac{Z_1^2}{2!} \, e^{2 \beta \mu} ~+~ ...
	~~=~~ \exp(Z_1 e^{\beta \mu}) 
\]
The expectation number of defects is
\[
\langle N \rangle 
~~=~~ \frac{\sum N_i \, e^{\beta(N_i \mu - E_i)}}{\mathcal{Z}}
~~=~~ \frac{1}{\beta \mathcal{Z}} \frac{\partial \mathcal{Z}}{\partial \mu}
~~=~~ \frac1\beta \, \frac{\partial \ln(\mathcal{Z})}{\partial \mu}
~~=~~ Z_1 \, e^{\beta \mu}
\] 

Substituting \eqref{eq:3d-partion-function} and $ \Delta H = \mu $ gives equation \eqDensity{} in the text 
\[
	\frac{\langle N \rangle}{N_0} ~~=~~ 3
	\left (\frac{2 \pi m^* s^2}{h^2 \beta}\right )^\frac12 
	~ e^{-\beta \Delta H}
\]

We now consider small terms which we have neglected in \eqDensity{},
and possible sources of systematic error in our comparison of the activation energy of
interstitial atoms and rotons (figure \figActivationTemperature{} inset).

An interstitial defect is resonant.
We have already counted its zero point energy,
since the chemical potential $ \mu $
is the energy required to create it
at low temperature, which
includes zero point energy,
but we have not counted its excited states.
Taking the lowest energy spherical harmonics, 
there are three orientations of resonance
which may be excited at higher temperature, whose frequency we estimate (below) to
be $ f \approx f_o = 1.6 \, 10^{11} $Hz
in superfluid helium at 1 atmosphere pressure.
This corresponds to a temperature of $ h f/K_B = 7.7K$. 
The partition function of an individual oscillator
(excluding zero point energy)
is given by $ (1-e^{-\beta h f})^{-1} $. Therefore the total partition function is modified by a factor of $ (1-e^{-\beta h f})^{-3}$.
Taking $ f \sim 1.6 \, 10^{11} $Hz, 
this gives a correction of 6.7\% at 2K, close to the highest temperatures plotted. 
This is small in relation to the agreement over
several orders of magnitude in the plot,
and we find it has little or no effect on our estimate of the
roton energy using a least squares fit.

We have also neglected perturbations to the existing phonon frequencies.
The speed of sound falls by approximately 4\% from 1K to 2K\cite{brooks1977calculated}.
We attribute this to the extra mass of the interstitial atoms.
Again, this is small in relation to the several orders of magnitude in the plot.

We used a constant value for
the effective mass of an interstitial atom, $ m^* = 0.165m $,
 when calculating the theoretical line
from \eqDensity{}.
This is the effective mass discussed in the text,
from our numerical calculation at atmospheric pressure and low temperature.
The effective mass of a roton,
which we associate with interstitial atoms,
reduces 
with pressure and temperature, both by of order 20\%\cite{brooks1977calculated}. 
The square root dependence on $ m^* $ in \eqDensity{} indicates a possible systematic error of order 10\%. 
This is also likely to have little effect on the activation energy for the same reason as above.

The roton activation energy in figure \figActivationTemperature{} (inset)
used the experimental values from from neutron scattering data at the lowest 
measured temperature, near 1.25K.
This slightly under-estimates the low temperature value.

In the main plot in figure \figActivationTemperature{}
we subtracted out the ordinary expansion of helium II by extrapolating from the expansion at low temperature,
where the concentration of interstitial atoms is negligible due to the
exponential decay.
This improved the match at low temperature,
but had little effect at higher temperatures where
the concentration of interstitial atoms
rises exponentially from \eqDensity{}.
At 15 atmospheres and above there was a reasonable fit to the expected $ T^3 $ dependence
of the thermal expansion coefficient at low temperature.
At lower pressures, 
there were also $ T^4 $ terms, which we speculate 
may be due to variations in the effective pressure 
associated with surface tension.
As a consequence the data at lower pressures is less accurate. Details can be seen in the spreadheet file.

We saw that interstitial atoms become aligned nematically when they are more concentrated, 
at higher temperature. This is likely to introduce
additional terms which we have not considered.
The rise in the graph near the transition temperature
may be associated with this.

\mySection{Further experimental evidence: rotons}

The following experiments also suggest that interstitial atoms have the
properties of rotons.
\begin{enumerate}[{b}1]

\item We would expect the effective mass of an interstitial defect 
to reduce with temperature, due to thermal 
agitation elongating it. The roton effective mass has a plateau below about 1K, and reduces sharply with temperature above this\cite{brooks1977calculated}.

\item In most other liquids, the speed of sound rises with temperature due to the reduction in density.  
In helium II, the mass of the interstitial atoms in \eqDensity{} raises the density and will reduce the speed of 
longitudinal sound.
It falls by approximately 4\% from 1K to 2K at 15 atmospheres pressure\cite{brooks1977calculated}. 

\item \label{b:second-sound} The mobile defects observed by Tucker and Wyatt\cite{tucker99antiparallel}
(which we associate with interstitial atoms)
transport heat energy,
giving helium II its high thermal conductivity.
If an interstitial atom moves through the lattice faster than the speed of transverse sound, we would
expect it to radiate an analogue of Cherenkov radiation, thereby limiting its maximum speed.
The velocity of heat transport (second sound) is limited to of order 20ms\textsuperscript{-1}, an order of magnitude less than the speed of longitudinal sound.

\item At warmer temperatures, 
interstitial atoms  
will exert a pressure due to their kinetic energy.
This is observed in the fountain effect,
a phenomenon also attributed to rotons\cite{feynman1953heliumabsolutezero,feynman1953heliumlambdatransition,feynman1954heliumtwofluid}.

\item An inhomogeneous electric field will trap
and stabilise interstitial atoms, 
which are denser and have 
stronger dielectric interactions. 
In 2007, Moroshkin, Hofer, Ulzega and Weis produced a dendritic solid by
melting the $ \gamma $ phase of \he4 with positively charged impurities sputtered onto it\cite{moroshkin2007impurity,moroshkin2008atomic,moroshkin2009positive}.
See the photograph in figure \ref{fig:snowballs}.
The dendrites are attracted to a cathode (to the right),
indicating they are positively charged.
They fall under gravity, indicating they are denser than helium II.

\begin{figure}[htb]
	\centering
	
	\includegraphics[width=.35\linewidth]{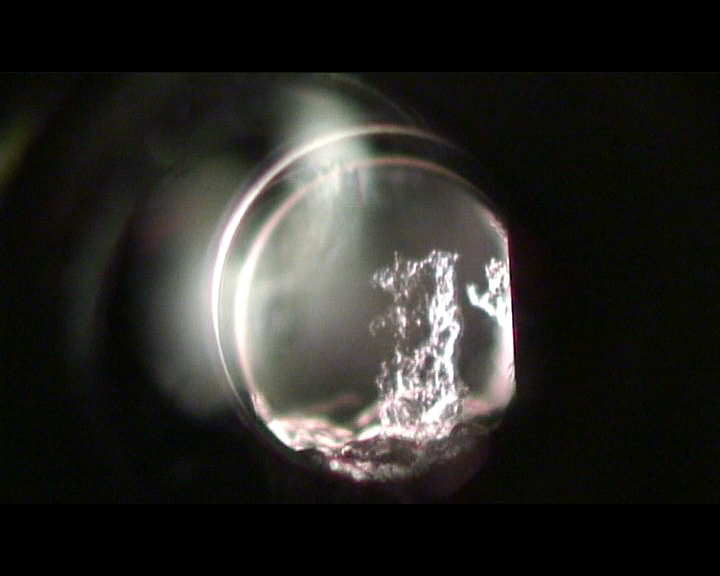}
	
	\caption{\em 
		Dendritic crystals in helium II, formed around charged impurities\cite{moroshkin2007impurity,moroshkin2008atomic,moroshkin2009positive},
		which we interpret as due to stabilised interstitial atoms.
		An alternative interpretation is that the
		photograph shows `frozen rotons'.
		Reproduced by kind permission of Peter Moroshkin.
		}
	\label{fig:snowballs}
\end{figure}



\end{enumerate}

\mySection{Further experimental evidence: fluidization}

The following additional observations relate to the fluidization of \he3, \he4 and colloids.

\begin{enumerate}[{c}1]


\newcommand{\CsolidifyHeating}{C4}
\item We noted (\CsolidifyHeating{}) that the factors driving the pretransition anomaly  in
$ \gamma $ helium
also apply elsewhere,
suggesting the existence of a narrow $ \gamma $-like pretransition phase very close to the melting curve.
There are experimental reports of a bcc-like phase
very close to the melting curve of \he4 near room temperature and 15GPa, 
but Frenkel calculated that an ordinary bcc structure would not be stable\cite{frenkel1986stability}.
He showed that quantum effects would not stabilise the phase, but
did not consider a pretransition $ \gamma $ phase,
which is stabilised by the entropy of the
interstitial atoms.
See later studies calling the experimental observations into question\cite{vos1990high}.

\item 
	We saw that helium II solidifies on heating near \singlequote{A} in figure \figPhaseDiagramHeFour{},
	and attributed it to the greater entropy of a pretransition $ \gamma $-like phase (see \CsolidifyHeating{}).
	Our explanation requires that the depression in the melting curve between 
	\singlequote{A} and \singlequote{B} in the figure cannot be wider than the anomaly itself.
	The anomaly in the specific heat capacity of $ \gamma $ helium is approximately 20mK wide\cite{hoffer1976thermodynamic},
	or, based on the slope of the curve, 20kPa,
	while the depression is only 1kPa.
		
\item The fluidized phase of \he3 also solidifies on heating
(see \singlequote{A} in figure \ref{fig:phase-diagram-he3}).
The solid in this region has the same x-ray characteristics as the $ \gamma $ phase of \he4,
indicating they have the same structure\cite{schuch1958structure,schuch1962structure}, namely a bcc lattice saturated with interstitial atoms, which may be favoured 
by the interactions between the magnetised nuclei in \he3.
The explanation for the solidification on heating in \CsolidifyHeating{} applies directly.
The magnitude of the depression in the melting curve is not limited by the width of a pretransition anomaly, and it is much larger than in \he4.  

\begin{figure}[htb]
	\centering
	\includegraphics[width=.8\linewidth]{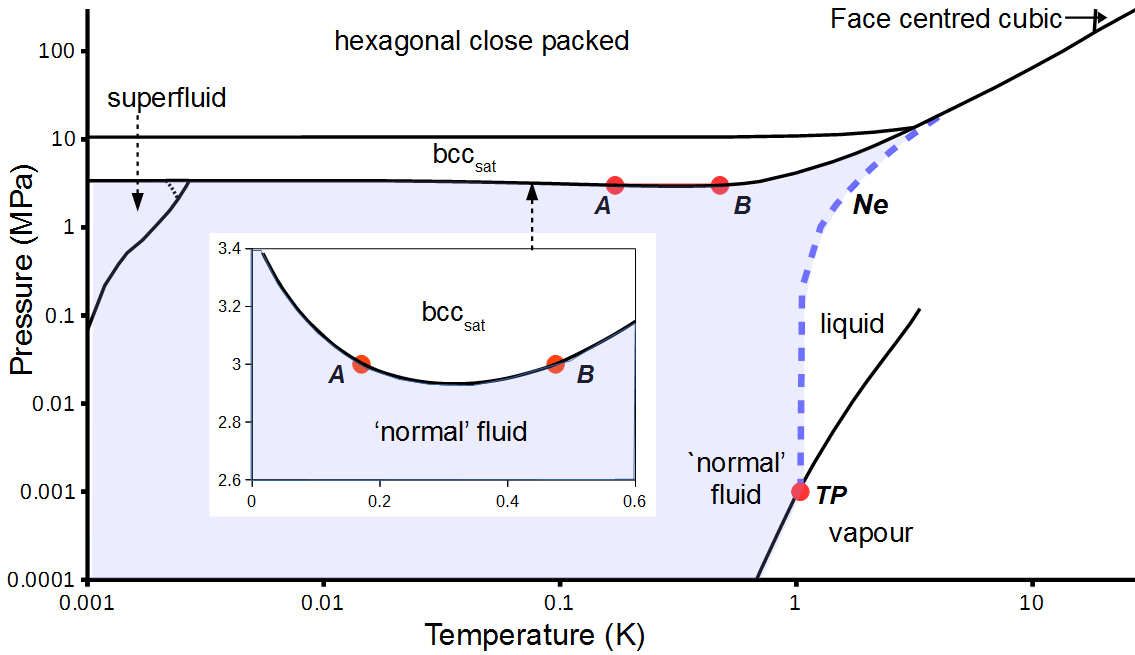}
	
	\caption{\em Phase diagram for \he3 on a logarithmic scale, with linear inset.
		The melting curve for neon is superposed 
		with its pressures and temperatures scaled so its
		triple point is at \singlequote{TP}\cite{scribner1969melting,young1975phase}. 
	}
	\label{fig:phase-diagram-he3}
\end{figure}

\item In 1959, Bernardes and Primakoff predicted that \he3
would solidify on heating
using different assumptions, 
namely that the solid is ordinary bcc crystals,
and it has more entropy than the liquid (which they assumed is a Fermi liquid) 
due to nuclear spins\cite{bernardes1959theory}.
However, it was later found that their model implies a spin ordering temperature much higher than observed\cite{scribner1969melting}.
They did not consider the entropy associated with interstitial defects,
an omission corrected above.

\item \label{SM:colloid-bcc} In 2012, Besseling, Hermes, Fortini, Dijkstra, Imhof and Blaaderen applied 
low amplitude oscillatory shear to a colloid
of spherical particles in the fluidized phase
near the solidification concentration, 
and observed the appearance of bcc-like order,
both in numerical simulation and in  experiments\cite{besseling2012oscillatory}.
The structure was body centred tetragonal (i.e. body centred cubic, slightly elongated perpendicular to the shear planes)
distorted into a hexagonal structure
at the extremes of the shear.	
We expect the onset of fluidization in this structure 
to be similar to our proposals for helium II.
One possibility is that this has already happened:
the bcc order in the colloidal fluid has been damaged by the flow,
and the gentle oscillation in this experiment
has helped to heal the damage to
reveal the underlying bcc-like order.

\end{enumerate}

\mySection{One-dimensional lattice equation of motion}

The equation of motion for a one-dimensional line of atoms is well known.
Suppose that atoms of mass $m$ are
distance $d$ apart, with an elastic constant $C_s$ between neighbours.
If the $n$'th atom is displaced by $U_n(t)$ then 
the force on it from its nearest neighbours will  be
\begin{equation}
	m \, \frac{d^2 U_n}{dt^2}
	~~=~~C_s(U_{n+1} +  U_{n-1} - 2 U_n)
	\label{eq:lattice-equation}
	\end{equation}
We can make a continuous approximation,
which is valid for long wavelength waves,
by defining a smooth function $\phi(x,t)$ 
so that $\phi(nd, t)=U_n(t)$.
Substituting the second-order Taylor expansion around $ x=0 $, namely 
	\[
		\phi(x, t) ~~=~~ \phi(0, t)
		 ~+~ x \,\frac{\partial \phi(0, t)}{\partial t}
		 ~+~ \frac12 x^2 \, \frac{\partial^2 \phi(0, t)}{\partial t^2}
	\]
	into \eqref{eq:lattice-equation} 
	gives 
	\[
		\frac{\partial^2 \phi}{\partial t^2} \, -\,  c_s^2 \, \frac{\partial^2 \phi}{\partial x^2} ~~ = ~~ 0
		\]
	which is the standard wave equation where the speed of sound is $c_s=d \sqrt{C_s/m}$.
	This describes low frequency sound waves in the lattice.

	There is another continuous approximation which is 
	valid near the maximum frequency of waves,
	when adjacent atoms oscillate almost antiphase with one another.
	We again define a smooth function $ \phi(x, t) $ so that $\phi(nd, t)= (-1)^n U_n(t)$.
	The same Taylor expansion gives
\begin{equation}\label{eq:optical-waves}
	\frac{\partial^2 \phi}{\partial t^2} \, + \,  c_s^2 \, \frac{\partial^2 \phi}{\partial x^2} ~~ = ~~ -\omega_o^2 \phi 
\end{equation}	
where $ \omega_o ^2= 4 C_s/m$ or $\omega_o =  2 c_s/d $. 

	We can describe the propagating waves just below the maximum frequency
	by substituting a solution of form 
	$ \phi \propto \cos(kx - \omega t) $.
	This gives the dispersion relation
	\begin{equation}\label{eq:dispersion}
		\omega^2~~=~~\omega_o^2 ~-~ c_s^2 k^2	
	\end{equation}		
	It follows immediately from \eqref{eq:dispersion}
	that propagating waves
	do not exist at angular frequencies above $ \omega_o $. 

	We can estimate $ \omega_o $ in \he4
at 1 atmosphere pressure 
for a bcc crystal structure where $ d $ is the side of a primitive cell and using the data reported in Brooks for helium II at atmospheric pressure and low temperature\cite{brooks1977calculated}, giving

\begin{center}
	\spacing{1}
	\begin{tabular}{llll}
		\\	$d$ & & $4.49 \, 10^{-10}$ & m
		\\	$c_s$ &  & 225 & m s$ ^{-1} $
		\\	$\omega_o$ & $2 c_s/d$ &  1.0\,10$ ^{12} $ & s$ ^{-1} $
		\\ $ f_o $ & $ \omega_o/(2 \pi) $ & 1.6\,10$ ^{11} $ & Hz
	\end{tabular}
	\spacing{2}
\end{center}

where $ f_o $ is the maximum frequency of propagating waves. This is a lower estimate for $ f_o $,
based on assuming $ d $ is the length of a bcc cell. An upper estimate is a factor $ \sqrt{3}/2 $
larger, based on the shortest distance between atoms.

\mySection{Resonant interstitial atoms}

Near an interstitial atom,
in a one-dimensional idealisation,
the equation of motion \eqref{eq:optical-waves}
has a localised solution 
in which the atoms oscillate almost antiphase.

In this solution, the interstitial atom itself  
is stationary at $ x=0 $ and it provides the boundary condition for the solution, which is
\[
	\phi ~~=~~\cos(\omega t) \, e^{-\mu |x|}
\]
This obeys \eqref{eq:optical-waves} when 
\[
\omega^2 ~~=~~\omega_o^2 + c_s^2 \mu^2
\]
Expressed in terms of the atomic displacements,
this solution is
\begin{equation}\label{eq:stationary-resonant-defect}
	U_n ~~\approx~~ (-1)^i \, \cos(\omega t) \, e^{-\mu |x_n|} 
\end{equation}
where the displacements $ U_n $ are mirror images in the origin, $ U_{-i} = - U_{i} $,
and the interstitial atom is stationary, $ U_0 = 0 $.

The above solution was for a stationary defect.
When the defect advances at velocity $ v $,
the corresponding solution is
	\[
		\phi ~~=~~
		e^{\nu t - \mu x }
		\, \cos(kx - \omega t)
	\]	
which obeys \eqref{eq:optical-waves} when
	\begin{equation}\label{eq:dipsersion-evanescent-waves}
		(\nu - i \omega)^2 ~+~ c_s^2(i k - \mu)^2~~=~~-\omega_o^2
	\end{equation}
and writing the atomic displacements explicitly gives the equation used in the text 
\[
	U_n ~~ \approx ~~
	(-1)^i
	~ e^{\nu t -\mu x_n}
	~ \cos(k x_n - \omega t)
\]

	The amplitude $ e^{\nu t - \mu x} $ advances with the defect at velocity $ v = \nu/\mu $
	since its value remains constant when $ x = (\nu/\mu)t + $ constant.
	Equating the imaginary parts of \eqref{eq:dipsersion-evanescent-waves} gives an the velocity
\begin{equation}\label{eq:defect-velocity}
		v ~~=~~ \frac{\nu}{\mu}~~=~~-\frac{c_s^2 k}{\omega}
\end{equation}
Equating the real part of \eqref{eq:dipsersion-evanescent-waves} gives
the dispersion relation
\[
	\nu^2 - \omega^2 + c_s^2(\mu^2-k^2) ~~=~~ -\omega_o^2	
\]
from which we can check that the group velocity is the same as the velocity of the defect 
\[
	v_g 
	~~=~~ \frac{\partial \omega}{\partial k}
	~~=~~ -\frac{c_s^2 k}{\omega} 
	~~=~~ v
\]
%
%
	
	The wavelength of this solution is $ \lambda = 2 \pi/|k| $. 
	Substituting into \eqref{eq:defect-velocity}
	and approximating 
	$ \omega = 2 \pi f \approx \omega_o $
	gives an approximate relationship between the
	momentum of a pair of interstitial atoms $ p $ and the wavelength of the waves
\[
		p
		~~=~~2 m^*v
		~~ \approx ~~\frac{2 m^* c_s^2}{f \lambda} ~~=~~\frac{h_s}{\lambda}
\]
where $ h_s = 2 m^* c_s^2/f$.
Approximating $ f \approx f_o = 2 c_s/d $ gives  $ h_s = 2 \pi m^* c_s d $.
This `acoustic Planck constant' is calculated in the spreadsheet densityhelium.ods and the results displayed figure \figPlanckConstant{}.
It uses data from Brooks\cite{brooks1977calculated} -- 
namely, the speed of sound, the effective mass of a roton
from neutron scattering measurements,
and the inter-atomic distance calculated from the density on a bcc arrangement of atoms in the (200) and (111) directions\cite{brooks1977calculated}.

\mySection{Extension to thee dimensions}

We have described the resonances of an interstitial atom using
a one-dimensional simplification. 
In three dimensions, the atoms in adjacent rows will be displaced,
and there are also solutions where the displacements are not parallel to the 
direction of motion.
We outline these extensions in turn.

One extension is based on a perturbation 
of the 
one-dimensional solution \eqref{eq:stationary-resonant-defect}
for the atoms in the central row of the defect.
This solution is perturbed because the displacements disturb the atoms in the adjacent rows due to transverse strains.
The associated forces are much smaller than for longitudinal strains,
and so we expect the perturbation to decay rapidly with distance from the central line.
This suggests the perturbation is a small effect.

There is some experimental support for this approach.
If the velocity of the defect exceeds the speed of
transverse sound in the crystal,
then we would expect it to lose energy to an
analogue of Cerenkov radiation.
The maximum velocity of a defect is indeed
significantly less than the speed of longitudinal sound (see b\ref{b:second-sound} above).

An alternative perspective is to note that a displaced atom exerts forces on the atoms near it,
not just those in the one-dimensional line.
In this idealization we neglect the bcc lattice entirely,
and it is necessary to extend the continuous approximation to the
equation of motion \eqref{eq:optical-waves} to
three dimensions.
The solutions are likely to involve spherical Bessel functions like those for an unbaffled
loudspeaker in the open air.

There is also another class of solutions in which 
the displacements are perpendicular to the direction of motion of the interstitial atom.
Suppose the indexes of the atoms are $ (i,j,k) $,
corresponding to the ($ x, y, z $) directions,
so their coordinates  
are $ (x_i,y_j,z_k) $. 
There is an approximate solution in which
the displacements  
of the atoms in the $ (x,y) $ plane 
are parallel to the $ y $ direction and have magnitude

\[
	U_{ijk}^{(y)} ~~\approx~~ (-1)^j \, 
 	e^{-\mu |y_j|} 
 	\cos(k x_i - \omega t)
\]

It is easily verified, using the perturbation approach described above,
that the dispersion relation of this solution is in approximately the same form as that for
a relativistic particle, namely
\[
	\omega^2 = \omega_o^2 + c_t^2 k^2
\] 
where $ c_t $ is the speed of transverse sound.

In the above solutions, the motion of adjacent atoms is approximately antiphase in one direction and approximately in phase in the other two directions. 
To complete the picture, there are also more complicated  solutions in which the 
motion is antiphase in two and three directions. 

\mySection{Schematic of the isotropic and nematic arrangements}

Figure \ref{fig:nematic-isotropic} is a schematic illustration of 
the isotropic and nematic
domains of the rod-like interstitial defects in helium II,
which are locally aligned with the bcc lattice.
The illustration is idealised. We saw from neutron scattering data (\CdamageToLattice{}) 
that the lattice loses correlation on distance
scales larger than the length of a defect;
this loss of correlation is not illustrated.

\begin{figure}[htb]
	\centering
	\includegraphics[width=.5\linewidth]{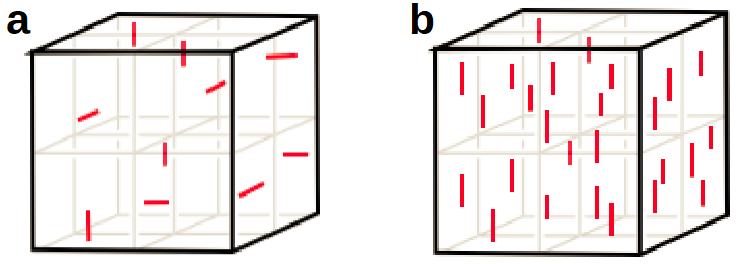}
	
	\caption{\em Schematic illustration of (a) isotropic and (b) nematic domains
		of rod-like interstitial defects in equilibrium in helium. 
		Distortions to the lattice,
		due to the fluidization, are not shown.
	}
	\label{fig:nematic-isotropic}
\end{figure}

\mySection{\large Critical comments}
\label{sec:critical-comments}

Experts in 
solid helium, quantum fluids, colloids
and soft matter have kindly offered 
critical comments which have greatly helped the manuscript.
Some observations arising are listed below.

\begin{enumerate}[Q1]
	\item \textbf{London repudiated his 1936 model 
		of helium II when he showed it is a quantum fluid.
		The manuscript fails to acknowledge this.}
	
	Contrary to some reports,
	London did not repudiate his original 1936 paper, which 
	described  the correlations among helium atoms in ordinary
	geometrical terms\cite{london1936condensed}.
	The stated aim of his 1938 paper\cite{london1938lambda} was to reject a proposal 
	by Fr\"ohlich that helium II
	has a diamond lattice\cite{frohlich1937lambda},
	and then to  ``direct attention to
	an entirely different interpretation''
	in momentum space.
	Representations in physical and momentum space are often complementary, and the manuscript attempts to show this in detail.
	Afterwards, London continued to advocate his 1936 description.
	In particular, in 1939 he emphasized that the rheology
	``depends essentially'' on volume\cite{london1939state}.
	
	\item \textbf{Helium II is a liquid.
		Liquids have insufficient positional order for
		the concept of an interstitial atom to be defined.}
	
	There is no evidence that
	the atoms in helium II are disordered 
	like in an ordinary liquid.
	London showed in 1936 that cold helium II has negligible entropy\cite{london1936condensed},
	and concluded that the atoms must be ordered positionally. 
	The manuscript describes the flow mechanism,
	which is similar to supersolid flow\cite{leggett1970can,andreev1969quantum}.	
	
	\item \textbf{The flowing phase of a colloid
		can be seen in a microscope, 
		but the predicted bcc-like order is not observed.}
	
	To the contrary, Besseling {\em et al} recently 
	photographed the appearance of bcc-like order
	in a colloid
	of spherical particles in the fluidized phase
	near the solidification concentration\cite{besseling2012oscillatory}.
	In their experiment they applied 
	low amplitude oscillatory shear to
	the phase. 
	There are a number of possible interpretations,
	but we suggest that the gentle oscillation
	accelerates the approach to equilibrium.
	See c\ref{SM:colloid-bcc} (in the supplementary material below)
	for further discussion.
	
	~~~~ In the absence of such stimulation,
	colloids approach equilibrium extremely slowly;
	for example, they typically take days to
	settle\cite{pusey2009hard}.
	Even after a long time they do not reach equilibrium on Earth, 
	as evidenced by the fact that they behave 
	differently on the space shuttle\cite{zhu1997crystallization}.
	This can be understood by noting that
	the arrangement of particles is very weak mechanically,
	particularly in the flowing phase,
	so that vibrations, gravity and convection
	impede or prevent the approach to equilibrium.
	
	~~~~ Helium approaches equilibrium faster than a colloid, due to the smaller scale, but
	nevertheless may not reach it on Earth
	since it appears to behave differently in space shuttle experiments\cite{kleinert2000theory}.

	\item \textbf{The $ \gamma $ phase of \he4lium has been classified body centred cubic for many years. 
		If this classification were mistaken, 
		as claimed,
		then it would have been discovered
		by specialists in the field by now.}   
	
	Specialists working on $ \gamma $ helium
	told us that the 7\% discrepancy in molar volume (figure \refFigMolarVolume) is well known.
	It is not the subject of active study because it is believed to have been
	explained historically.
	We traced this supposed explanation to three papers, 
	by Schuch {\em et al} in 1958, 1961 and 1962\cite{schuch1961bond,schuch1962structure,schuch1958structure} and noticed that the 1961 and 1962 papers 
	contradict each other, as discussed in the text. 
	These papers also reported unexplained anomalies in the x-ray patterns and crystal size
	which led us directly to the structure 
	in figure \refFigBccToFcc.
	
	\item  \textbf{Does the manuscript make observable predictions 
		that differ from the conventional model?}
	
	Yes. The manuscript predicts that the lowest energy excitations
	in helium II (other than acoustic phonons) are interstitial atoms, which 
	have the gravimetric mass $ m $ of a helium atom
	and effective mass
	(from the relationship between velocity
	and kinetic energy) of $ m^* \approx 0.165m $ (see \BeffectiveMass).
	Both are close to the observed values
	-- the effective mass is known from neutron scattering experiments\cite{brooks1977calculated} 
	and the mass can be inferred from its effect on density (see figure \refFigActivationTemperature).
	
	~~~ According to the conventional model, the lowest energy excitations are
	rotons. The theory does not predict their effective mass, 
	but Feynman thought a roton's kinetic energy is
	primarily associated with circular flow patterns\cite{feynman1954heliumtwofluid},
	from which it follows that their gravimetric mass is much smaller
	than $ m^* $,
	contrary to the negative thermal expansion measurements plotted in figure \refFigActivationTemperature.
	
	\item \textbf{The manuscript claims that the flow
		in helium II is carried by interstitial helium atoms,
		which are mobile.
		But at low temperature
		their concentration is vanishingly small,
		as shown in figure \figActivationTemperature}
	
	Figure \figActivationTemperature{} is the concentration of interstitial atoms in thermal equilibrium,
	but flow is not an equilibrium phenomenon.
	Interstitial atoms can be formed near surface imperfections
	by mechanical, rather then thermal, energy,
	as described in the text.
	They are metastable, since helium atoms cannot be destroyed, and continue to carry the flow for a considerable time.
	
	\newcommand{\eqWavelength}{(4)}

	\item \textbf{The manuscript 
		describes helium using classical equations of motion. 
		But helium is a quantum fluid.
		Is this an attempt to dethrone quantum mechanics?}
	
	No. Textbooks on quantum fluids typically begin with the wavelength postulate $ \lambda = h/p $ where $ p $ is the momentum\cite{leggett2006quantum}.
	The wavelength in \eqWavelength{},
	$ \lambda = h_s/p $,
	was derived from ordinary classical equations of motion
	in the same way as for the experiments on bouncing droplets\cite{bush2015new,bradyanderson2014why,filoux2015strings}.
	It was a surprise to the authors that the parameter
	$ h_s $ is empirically the same as Planck's constant
	to experimental accuracy at all pressures (figure \refFigPlanckConstant).
	This suggests that quantum processes ultimately underly the
	result. 
	In particular, the forces between helium atoms
	are quantum mechanical in origin. 
	We suggest this is an area for further research.
	
	\item \textbf{Liquid \he4 is a Bose-Einstein condensate and 
		liquid \he3 is a Fermi liquid.
		Does the manuscript argue otherwise?}
	
	Not necessarily. `Bose-Einstein condensate'
	and `Fermi liquid' are descriptions in momentum space
	whereas the manuscript 
	discusses a geometrical representation, in physical space.
	
	~~~~ However, there is a difference between the models
	which might, in future, be measured experimentally.
	The wavelength \eqWavelength{}
	depends on the momentum of a pair of interstitial atoms,
	based on their effective mass,
	while London's 1938 model of a Bose-Einstein condensate involves the gravimetric 
	mass of a single helium atom.
	This is approximately a factor of 3 larger. 
	We are not aware of any experiments
	to date which are capable of distinguishing between them.
	Compare the corresponding electron mass in superconductors, according to the conventional model\cite{brady1982correction}.
	
	\item \textbf{The phase diagram of \he4 has 
		the `$ \lambda $' line, below which superflow is observed
		on a macroscopic scale, and the 
		line of maximum density which is very close to it. Which 
		of these is the actual thermodynamic transition?}

	The line of maximum density ($ \rho_{max} $ in figure
	\figPhaseDiagramHeFour{} inset)
	is the thermodynamic transition.
	To the right of it, the interstitial atoms
	are arranged nematically,
	and to the left there is a biphase of
	nematic and isotropic domains.
	Ordinary liquid crystals are similar\cite{onsager1949effects}.
	
	~~~~ The nematic phase is an efficient packing arrangement
	which lacks room for more interstitial atoms,
	so it is difficult or impossible for their concentration 
	to rise with temperature in this region.
	This accounts for the sudden fall in the specific heat capacity
	and the thermal expansion coefficient reversing sign.
	The isotropic domains (which we associate with superfluid phenomena, as discussed in the text)
	are quenched to the right of $ \rho_{max} $ and
	superfluid
	phenomena are quenched accordingly.
	
	%
	%
	%
	%
	~~~~ Cooling the substance below $ \rho_{max} $, the 
	next event occurs at the line $ \lambda $,
	where the isotropic domains join up
	and superflow is possible over a macroscopic sample.
	This is a geometrical
	phenomenon rather than an ordinary thermodynamic transition.
	We would expect superfluid behaviour (other than macroscopic flow)
	to persist between the $ \lambda $ and $ \rho_{max} $ lines.
	This is observed and called 
	the superfluid fluctuation regime.

	~~~~ The approximately exponential
	rise in the specific heat capacity from of order 1K to the
	transition temperature can be understood in the same way as in the conventional
	theory, as due to the exponential rise in the number of
	excitations (equation \eqDensity{}).
	Very near the transition,
	there is a spike in the specific heat capacity.
	We conjecture that this is associated with nonlinear effects due to the vanishingly small size of the isotropic domains,
	an area for further research. 
	
\end{enumerate}

\fi

\bibliography{book}

\begin{thebibliography}{10}
\expandafter\ifx\csname url\endcsname\relax
  \def\url#1{\texttt{#1}}\fi
\expandafter\ifx\csname urlprefix\endcsname\relax\def\urlprefix{URL }\fi
\providecommand{\bibinfo}[2]{#2}
\providecommand{\eprint}[2][]{\url{#2}}

\bibitem{young1975phase}
\bibinfo{author}{Young, D.}
\newblock \emph{\bibinfo{title}{The phase diagrams of the elements}}
  (\bibinfo{publisher}{Lawrence Livermore Laboratory}, \bibinfo{year}{1975}).

\bibitem{donnelly1998observed}
\bibinfo{author}{Donnelly, R.} \& \bibinfo{author}{Barenghi, C.}
\newblock \bibinfo{title}{The observed properties of liquid helium at the
  saturated vapor pressure}.
\newblock \emph{\bibinfo{journal}{Journal of Physical and Chemical Reference
  Data}} \textbf{\bibinfo{volume}{27}}, \bibinfo{pages}{1217--1274}
  (\bibinfo{year}{1998}).

\bibitem{brooks1977calculated}
\bibinfo{author}{Brooks, J.} \& \bibinfo{author}{Donnelly, R.}
\newblock \bibinfo{title}{The calculated thermodynamic properties of superfluid
  helium-4}.
\newblock \emph{\bibinfo{journal}{Journal of Physical and Chemical Reference
  Data}} \textbf{\bibinfo{volume}{6}}, \bibinfo{pages}{51--104}
  (\bibinfo{year}{1977}).

\bibitem{hoffer1976thermodynamic}
\bibinfo{author}{Hoffer, J.}, \bibinfo{author}{Gardner, W.},
  \bibinfo{author}{Waterfield, C.} \& \bibinfo{author}{Phillips, N.}
\newblock \bibinfo{title}{Thermodynamic properties of {$ ^4 $He. II. T}he bcc
  phase and the {P-T and VT} phase diagrams below 2 {K}}.
\newblock \emph{\bibinfo{journal}{Journal of Low Temperature Physics}}
  \textbf{\bibinfo{volume}{23}}, \bibinfo{pages}{63--102}
  (\bibinfo{year}{1976}).

\bibitem{schuch1962structure}
\bibinfo{author}{Schuch, A.} \& \bibinfo{author}{Mills, R.}
\newblock \bibinfo{title}{Structure of the $\gamma$ form of {Solid He} 4}.
\newblock \emph{\bibinfo{journal}{Physical Review Letters}}
  \textbf{\bibinfo{volume}{8}}, \bibinfo{pages}{469} (\bibinfo{year}{1962}).

\bibitem{london1936condensed}
\bibinfo{author}{London, F.}
\newblock \bibinfo{title}{On condensed helium at absolute zero}.
\newblock \emph{\bibinfo{journal}{Proceedings of the Royal Society of London.
  Series A, Mathematical and Physical Sciences}}
  \textbf{\bibinfo{volume}{153}}, \bibinfo{pages}{576--583}
  (\bibinfo{year}{1936}).

\bibitem{frohlich1937lambda}
\bibinfo{author}{Fr\"ohlich, H.}
\newblock \bibinfo{title}{Zur theorie des $ \lambda $-punktes des heliums}.
\newblock \emph{\bibinfo{journal}{Physica}} \textbf{\bibinfo{volume}{IV}},
  \bibinfo{pages}{639--644} (\bibinfo{year}{1937}).

\bibitem{london1939state}
\bibinfo{author}{London, F.}
\newblock \bibinfo{title}{The state of liquid helium near absolute zero.}
\newblock \emph{\bibinfo{journal}{Journal of Physical Chemistry}}
  \textbf{\bibinfo{volume}{43}}, \bibinfo{pages}{49--69}
  (\bibinfo{year}{1939}).

\bibitem{andreev1969quantum}
\bibinfo{author}{Andreev, A.} \& \bibinfo{author}{Lifshitz, I.}
\newblock \bibinfo{title}{Quantum theory of defects in crystals}.
\newblock \emph{\bibinfo{journal}{Sov. Phys. JETP}}
  \textbf{\bibinfo{volume}{29}}, \bibinfo{pages}{1107--1113}
  (\bibinfo{year}{1969}).

\bibitem{leggett1970can}
\bibinfo{author}{Leggett, A.}
\newblock \bibinfo{title}{Can a solid be `superfluid'?}
\newblock \emph{\bibinfo{journal}{Physical Review Letters}}
  \textbf{\bibinfo{volume}{25}}, \bibinfo{pages}{1543} (\bibinfo{year}{1970}).

\bibitem{pusey1986phase}
\bibinfo{author}{Pusey, P.~N.} \& \bibinfo{author}{Van~Megen, W.}
\newblock \bibinfo{title}{Phase behaviour of concentrated suspensions of nearly
  hard colloidal spheres}.
\newblock \emph{\bibinfo{journal}{Nature}} \textbf{\bibinfo{volume}{320}},
  \bibinfo{pages}{340--342} (\bibinfo{year}{1986}).

\bibitem{pusey2009hard}
\bibinfo{author}{Pusey, P.} \emph{et~al.}
\newblock \bibinfo{title}{Hard spheres: crystallization and glass formation}.
\newblock \emph{\bibinfo{journal}{Philosophical Transactions of the Royal
  Society of London A: Mathematical, Physical and Engineering Sciences}}
  \textbf{\bibinfo{volume}{367}}, \bibinfo{pages}{4993--5011}
  (\bibinfo{year}{2009}).

\bibitem{london1938lambda}
\bibinfo{author}{London, F.}
\newblock \bibinfo{title}{The $ \lambda $-phenomenon of liquid helium and the
  {Bose-Einstein} degeneracy}.
\newblock \emph{\bibinfo{journal}{Nature}} \textbf{\bibinfo{volume}{141}},
  \bibinfo{pages}{643--644} (\bibinfo{year}{1938}).

\bibitem{schuch1961bond}
\bibinfo{author}{Schuch, A.}
\newblock \bibinfo{title}{Bond-lengths in solid helium}.
\newblock \emph{\bibinfo{journal}{Nature}} \textbf{\bibinfo{volume}{191}},
  \bibinfo{pages}{591} (\bibinfo{year}{1961}).

\bibitem{zhang2005experimental}
\bibinfo{author}{Zhang, J.} \emph{et~al.}
\newblock \bibinfo{title}{Experimental constraints on the phase diagram of
  elemental zirconium}.
\newblock \emph{\bibinfo{journal}{Journal of Physics and Chemistry of Solids}}
  \textbf{\bibinfo{volume}{66}}, \bibinfo{pages}{1213--1219}
  (\bibinfo{year}{2005}).

\bibitem{pauling1947atomic}
\bibinfo{author}{Pauling, L.}
\newblock \bibinfo{title}{Atomic radii and interatomic distances in metals}.
\newblock \emph{\bibinfo{journal}{Journal of the American Chemical Society}}
  \textbf{\bibinfo{volume}{69}}, \bibinfo{pages}{542--553}
  (\bibinfo{year}{1947}).

\bibitem{schuch1958structure}
\bibinfo{author}{Schuch, A.}, \bibinfo{author}{Grilly, E.} \&
  \bibinfo{author}{Mills, R.}
\newblock \bibinfo{title}{Structure of the $\alpha$ and $\beta$ forms of solid
  {He 3}}.
\newblock \emph{\bibinfo{journal}{Physical Review}}
  \textbf{\bibinfo{volume}{110}}, \bibinfo{pages}{775} (\bibinfo{year}{1958}).

\bibitem{moroshkin2008atomic}
\bibinfo{author}{Moroshkin, P.}, \bibinfo{author}{Hofer, A.} \&
  \bibinfo{author}{Weis, A.}
\newblock \bibinfo{title}{Atomic and molecular defects in solid {$ ^4 $He}}.
\newblock \emph{\bibinfo{journal}{Physics Reports}}
  \textbf{\bibinfo{volume}{469}}, \bibinfo{pages}{1--57}
  (\bibinfo{year}{2008}).

\bibitem{feynman1954heliumtwofluid}
\bibinfo{author}{Feynman, R.}
\newblock \bibinfo{title}{Atomic theory of the two-fluid model of liquid
  helium}.
\newblock \emph{\bibinfo{journal}{Physical Review}}
  \textbf{\bibinfo{volume}{94}}, \bibinfo{pages}{262} (\bibinfo{year}{1954}).

\bibitem{tucker99antiparallel}
\bibinfo{author}{Tucker, M.} \& \bibinfo{author}{Wyatt, A.}
\newblock \bibinfo{title}{Direct evidence for {R}$^-$ rotons having
  antiparallel momentum and velocity}.
\newblock \emph{\bibinfo{journal}{Science}} \textbf{\bibinfo{volume}{283}},
  \bibinfo{pages}{1150--1152} (\bibinfo{year}{1999}).

\bibitem{landau1947theory}
\bibinfo{author}{Landau, L.}
\newblock \bibinfo{title}{On the theory of superfluidity of helium {II}}.
\newblock \emph{\bibinfo{journal}{J Physics, Moscow}}
  \textbf{\bibinfo{volume}{11}}, \bibinfo{pages}{91--92}
  (\bibinfo{year}{1947}).

\bibitem{allum1976rotonpair}
\bibinfo{author}{Allum, D.~R.}, \bibinfo{author}{Bowley, R.~M.} \&
  \bibinfo{author}{McClintock, P. V.~E.}
\newblock \bibinfo{title}{Evidence for roton pair creation in superfluid
  $^4${He}}.
\newblock \emph{\bibinfo{journal}{Physical Review Letters}}
  \textbf{\bibinfo{volume}{36}} (\bibinfo{year}{1976}).

\bibitem{kim2004probable}
\bibinfo{author}{Kim, E.} \& \bibinfo{author}{Chan, M.}
\newblock \bibinfo{title}{Probable observation of a supersolid helium phase}.
\newblock \emph{\bibinfo{journal}{Nature}} \textbf{\bibinfo{volume}{427}},
  \bibinfo{pages}{225--227} (\bibinfo{year}{2004}).

\bibitem{maris2012effect}
\bibinfo{author}{Maris, H.}
\newblock \bibinfo{title}{Effect of elasticity on torsional oscillator
  experiments probing the possible supersolidity of helium}.
\newblock \emph{\bibinfo{journal}{Physical Review B}}
  \textbf{\bibinfo{volume}{86}}, \bibinfo{pages}{020502}
  (\bibinfo{year}{2012}).

\bibitem{balibar2016dislocations}
\bibinfo{author}{Balibar, S.} \emph{et~al.}
\newblock \bibinfo{title}{Dislocations in a quantum crystal: Solid helium: A
  model and an exception}.
\newblock \emph{\bibinfo{journal}{Comptes Rendus Physique}}
  \textbf{\bibinfo{volume}{17}}, \bibinfo{pages}{264--275}
  (\bibinfo{year}{2016}).

\bibitem{li2016observation}
\bibinfo{author}{Li, J.} \emph{et~al.}
\newblock \bibinfo{title}{Observation of the supersolid stripe phase in
  spin-orbit coupled bose-einstein condensates}.
\newblock \emph{\bibinfo{journal}{arXiv preprint arXiv:1610.08194}}
  (\bibinfo{year}{2016}).

\bibitem{leonard2016supersolid}
\bibinfo{author}{L{\'e}onard, J.}, \bibinfo{author}{Morales, A.},
  \bibinfo{author}{Zupancic, P.}, \bibinfo{author}{Esslinger, T.} \&
  \bibinfo{author}{Donner, T.}
\newblock \bibinfo{title}{Supersolid formation in a quantum gas breaking
  continuous translational symmetry}.
\newblock \emph{\bibinfo{journal}{arXiv preprint arXiv:1609.09053}}
  (\bibinfo{year}{2016}).

\bibitem{haziot2013giant}
\bibinfo{author}{Haziot, A.}, \bibinfo{author}{Rojas, X.},
  \bibinfo{author}{Fefferman, A.~D.}, \bibinfo{author}{Beamish, J.~R.} \&
  \bibinfo{author}{Balibar, S.}
\newblock \bibinfo{title}{Giant plasticity of a quantum crystal}.
\newblock \emph{\bibinfo{journal}{Physical review letters}}
  \textbf{\bibinfo{volume}{110}}, \bibinfo{pages}{035301}
  (\bibinfo{year}{2013}).

\bibitem{caupin2008static}
\bibinfo{author}{Caupin, F.}, \bibinfo{author}{Boronat, J.} \&
  \bibinfo{author}{Andersen, K.}
\newblock \bibinfo{title}{Static structure factor and static response function
  of superfluid helium 4: A comparative analysis}.
\newblock \emph{\bibinfo{journal}{Journal of Low Temperature Physics}}
  \textbf{\bibinfo{volume}{152}}, \bibinfo{pages}{108--121}
  (\bibinfo{year}{2008}).

\bibitem{zhu1997crystallization}
\bibinfo{author}{Zhu, J.} \emph{et~al.}
\newblock \bibinfo{title}{Crystallization of hard-sphere colloids in
  microgravity}.
\newblock \emph{\bibinfo{journal}{Nature}} \textbf{\bibinfo{volume}{387}},
  \bibinfo{pages}{883--885} (\bibinfo{year}{1997}).

\bibitem{donovan1971lattice}
\bibinfo{author}{Donovan, B.} \& \bibinfo{author}{Angress, J.}
\newblock \emph{\bibinfo{title}{Lattice vibrations}} (\bibinfo{year}{1971}).

\bibitem{fort2010path}
\bibinfo{author}{Fort, E.}, \bibinfo{author}{Eddi, A.},
  \bibinfo{author}{Boudaoud, A.}, \bibinfo{author}{Moukhtar, J.} \&
  \bibinfo{author}{Couder, Y.}
\newblock \bibinfo{title}{Path-memory induced quantization of classical
  orbits}.
\newblock \emph{\bibinfo{journal}{Proceedings of the National Academy of
  Sciences}} \textbf{\bibinfo{volume}{107}}, \bibinfo{pages}{17515--17520}
  (\bibinfo{year}{2010}).

\bibitem{couder2006single}
\bibinfo{author}{Couder, Y.} \& \bibinfo{author}{Fort, E.}
\newblock \bibinfo{title}{Single-particle diffraction and interference at a
  macroscopic scale}.
\newblock \emph{\bibinfo{journal}{Phy. Rev. Lett.}}
  \textbf{\bibinfo{volume}{97}}, \bibinfo{pages}{154101}
  (\bibinfo{year}{2006}).

\bibitem{eddi2009unpredictable}
\bibinfo{author}{Eddi, A.}, \bibinfo{author}{Fort, E.}, \bibinfo{author}{Moisy,
  F.} \& \bibinfo{author}{Couder, Y.}
\newblock \bibinfo{title}{Unpredictable tunneling of a classical wave-particle
  association}.
\newblock \emph{\bibinfo{journal}{Physical review letters}}
  \textbf{\bibinfo{volume}{102}}, \bibinfo{pages}{240401}
  (\bibinfo{year}{2009}).

\bibitem{bush2015new}
\bibinfo{author}{Bush, J.}
\newblock \bibinfo{title}{The new wave of pilot-wave theory}.
\newblock \emph{\bibinfo{journal}{Physics Today}}
  \textbf{\bibinfo{volume}{68}}, \bibinfo{pages}{47--53}
  (\bibinfo{year}{2015}).

\bibitem{bradyanderson2014why}
\bibinfo{author}{Brady, R.} \& \bibinfo{author}{Anderson, R.}
\newblock \bibinfo{title}{Why bouncing droplets are a pretty good model of
  quantum mechanics}.
\newblock \emph{\bibinfo{journal}{arXiv:1401.4356}}  (\bibinfo{year}{2014}).

\bibitem{acebron2005kuramoto}
\bibinfo{author}{Acebr{\'o}n, J.}, \bibinfo{author}{Bonilla, L.},
  \bibinfo{author}{Vicente, C.}, \bibinfo{author}{Ritort, F.} \&
  \bibinfo{author}{Spigler, R.}
\newblock \bibinfo{title}{The {Kuramoto} model: A simple paradigm for
  synchronization phenomena}.
\newblock \emph{\bibinfo{journal}{Reviews of Modern Physics}}
  \textbf{\bibinfo{volume}{77}}, \bibinfo{pages}{137} (\bibinfo{year}{2005}).

\bibitem{bennett2002huygens}
\bibinfo{author}{Bennett, M.}, \bibinfo{author}{Schatz, M.~F.},
  \bibinfo{author}{Rockwood, H.} \& \bibinfo{author}{Wiesenfeld, K.}
\newblock \bibinfo{title}{Huygens's clocks}.
\newblock \emph{\bibinfo{journal}{Proc. Roy. Soc. A}}
  \textbf{\bibinfo{volume}{458}}, \bibinfo{pages}{563--579}
  (\bibinfo{year}{2002}).

\bibitem{filoux2015strings}
\bibinfo{author}{Filoux, B.}, \bibinfo{author}{Hubert, M.} \&
  \bibinfo{author}{Vandewalle, N.}
\newblock \bibinfo{title}{Strings of droplets propelled by coherent waves}.
\newblock \emph{\bibinfo{journal}{Physical Review E}}
  \textbf{\bibinfo{volume}{92}}, \bibinfo{pages}{041004}
  (\bibinfo{year}{2015}).

\bibitem{leggett2006quantum}
\bibinfo{author}{Leggett, A.~J.}
\newblock \emph{\bibinfo{title}{Quantum liquids: Bose condensation and Cooper
  pairing in condensed-matter systems}} (\bibinfo{publisher}{Oxford University
  Press}, \bibinfo{year}{2006}).

\bibitem{bardeen1957theory}
\bibinfo{author}{Bardeen, J.}, \bibinfo{author}{Cooper, L.~N.} \&
  \bibinfo{author}{Schrieffer, J.~R.}
\newblock \bibinfo{title}{Theory of superconductivity}.
\newblock \emph{\bibinfo{journal}{Physical Review}}
  \textbf{\bibinfo{volume}{108}}, \bibinfo{pages}{1175} (\bibinfo{year}{1957}).

\bibitem{josephson1962coupled}
\bibinfo{author}{Josephson, B.}
\newblock \emph{\bibinfo{title}{The relativistic shift in the M\"ossbauer
  effect and coupled superconductors}} (\bibinfo{publisher}{Fellowship
  dissertation, Trinity College, Cambridge}, \bibinfo{year}{1962 \\
  {\scriptsize \url{www.dspace.cam.ac.uk/handle/1810/243916}}}).

\bibitem{panczyk1967evidence}
\bibinfo{author}{Panczyk, M.}, \bibinfo{author}{Scribner, R.},
  \bibinfo{author}{Straty, G.} \& \bibinfo{author}{Adams, E.}
\newblock \bibinfo{title}{Evidence of nuclear spin ordering in solid
  helium-three}.
\newblock \emph{\bibinfo{journal}{Physical Review Letters}}
  \textbf{\bibinfo{volume}{19}}, \bibinfo{pages}{1102} (\bibinfo{year}{1967}).

\bibitem{onsager1949effects}
\bibinfo{author}{Onsager, L.}
\newblock \bibinfo{title}{The effects of shape on the interaction of colloidal
  particles}.
\newblock \emph{\bibinfo{journal}{Annals of the New York Academy of Sciences}}
  \textbf{\bibinfo{volume}{51}}, \bibinfo{pages}{627--659}
  (\bibinfo{year}{1949}).

\bibitem{degennes1974liquidcrystals}
\bibinfo{author}{de~Gennes, P.}
\newblock \emph{\bibinfo{title}{The Physics of Liquid Crystals}}
  (\bibinfo{publisher}{Oxford}, \bibinfo{year}{1974}).

\bibitem{lekkerkerker2013liquid}
\bibinfo{author}{Lekkerkerker, H.} \& \bibinfo{author}{Vroege, G.}
\newblock \bibinfo{title}{Liquid crystal phase transitions in suspensions of
  mineral colloids: new life from old roots}.
\newblock \emph{\bibinfo{journal}{Phil. Trans. R. Soc. A}}
  \textbf{\bibinfo{volume}{371}}, \bibinfo{pages}{20120263}
  (\bibinfo{year}{2013}).

\bibitem{balibar2014superfluidity}
\bibinfo{author}{Balibar, S.}
\newblock \bibinfo{title}{Superfluidity: How quantum mechanics became visible}.
\newblock In \emph{\bibinfo{booktitle}{History of Artificial Cold, Scientific,
  Technological and Cultural Issues}}, \bibinfo{pages}{93--117}
  (\bibinfo{publisher}{Springer}, \bibinfo{year}{2014}).

\bibitem{kobyakov2014towards}
\bibinfo{author}{Kobyakov, D.} \& \bibinfo{author}{Pethick, C.}
\newblock \bibinfo{title}{Towards a metallurgy of neutron star crusts}.
\newblock \emph{\bibinfo{journal}{Physical review letters}}
  \textbf{\bibinfo{volume}{112}}, \bibinfo{pages}{112504}
  (\bibinfo{year}{2014}).

\bibitem{vocadlo2003possible}
\bibinfo{author}{Vo{\v{c}}adlo, L.} \emph{et~al.}
\newblock \bibinfo{title}{Possible thermal and chemical stabilization of
  body-centred-cubic iron in the earth's core}.
\newblock \emph{\bibinfo{journal}{Nature}} \textbf{\bibinfo{volume}{424}},
  \bibinfo{pages}{536--539} (\bibinfo{year}{2003}).

\bibitem{errea2016quantum}
\bibinfo{author}{Errea, I.} \emph{et~al.}
\newblock \bibinfo{title}{Quantum hydrogen-bond symmetrization in the
  superconducting hydrogen sulfide system}.
\newblock \emph{\bibinfo{journal}{Nature}} \textbf{\bibinfo{volume}{532}},
  \bibinfo{pages}{81} (\bibinfo{year}{2016}).

\bibitem{aziz1995ab}
\bibinfo{author}{Aziz, R.~A.}, \bibinfo{author}{Janzen, A.~R.} \&
  \bibinfo{author}{Moldover, M.~R.}
\newblock \bibinfo{title}{Ab initio calculations for helium: a standard for
  transport property measurements}.
\newblock \emph{\bibinfo{journal}{Physical review letters}}
  \textbf{\bibinfo{volume}{74}}, \bibinfo{pages}{1586} (\bibinfo{year}{1995}).

\bibitem{feynman1953heliumabsolutezero}
\bibinfo{author}{Feynman, R.}
\newblock \bibinfo{title}{Atomic theory of liquid helium near absolute zero}.
\newblock \emph{\bibinfo{journal}{Physical Review}}
  \textbf{\bibinfo{volume}{91}}, \bibinfo{pages}{1301} (\bibinfo{year}{1953}).

\bibitem{feynman1953heliumlambdatransition}
\bibinfo{author}{Feynman, R.}
\newblock \bibinfo{title}{Atomic theory of the $\lambda$ transition in helium}.
\newblock \emph{\bibinfo{journal}{Physical Review}}
  \textbf{\bibinfo{volume}{91}}, \bibinfo{pages}{1291} (\bibinfo{year}{1953}).

\bibitem{moroshkin2007impurity}
\bibinfo{author}{Moroshkin, P.}, \bibinfo{author}{Hofer, A.},
  \bibinfo{author}{Ulzega, S.} \& \bibinfo{author}{Weis, A.}
\newblock \bibinfo{title}{Impurity-stabilized solid {$ ^4 $He} below the
  solidification pressure of pure helium}.
\newblock \emph{\bibinfo{journal}{Nature Physics}}
  \textbf{\bibinfo{volume}{3}}, \bibinfo{pages}{786--789}
  (\bibinfo{year}{2007}).

\bibitem{moroshkin2009positive}
\bibinfo{author}{Moroshkin, P.}, \bibinfo{author}{Lebedev, V.} \&
  \bibinfo{author}{Weis, A.}
\newblock \bibinfo{title}{Positive ion induced solidification of he 4}.
\newblock \emph{\bibinfo{journal}{Physical review letters}}
  \textbf{\bibinfo{volume}{102}}, \bibinfo{pages}{115301}
  (\bibinfo{year}{2009}).

\bibitem{frenkel1986stability}
\bibinfo{author}{Frenkel, D.}
\newblock \bibinfo{title}{Stability of the high-pressure body-centered-cubic
  phase of helium}.
\newblock \emph{\bibinfo{journal}{Physical review letters}}
  \textbf{\bibinfo{volume}{56}}, \bibinfo{pages}{858} (\bibinfo{year}{1986}).

\bibitem{vos1990high}
\bibinfo{author}{Vos, W.~L.}, \bibinfo{author}{van Hinsberg, M.~G.} \&
  \bibinfo{author}{Schouten, J.~A.}
\newblock \bibinfo{title}{High-pressure triple point in helium: The melting
  line of helium up to 240 kbar}.
\newblock \emph{\bibinfo{journal}{Physical Review B}}
  \textbf{\bibinfo{volume}{42}}, \bibinfo{pages}{6106} (\bibinfo{year}{1990}).

\bibitem{scribner1969melting}
\bibinfo{author}{Scribner, R.}, \bibinfo{author}{Panczyk, M.} \&
  \bibinfo{author}{Adams, E.}
\newblock \bibinfo{title}{Melting curve and related thermodynamic properties of
  {$ ^3 $He below 1 K}}.
\newblock \emph{\bibinfo{journal}{Journal of Low Temperature Physics}}
  \textbf{\bibinfo{volume}{1}}, \bibinfo{pages}{313--340}
  (\bibinfo{year}{1969}).

\bibitem{bernardes1959theory}
\bibinfo{author}{Bernardes, N.} \& \bibinfo{author}{Primakoff, H.}
\newblock \bibinfo{title}{Theory of solid {He$ ^3 $}}.
\newblock \emph{\bibinfo{journal}{Physical Review Letters}}
  \textbf{\bibinfo{volume}{2}}, \bibinfo{pages}{290} (\bibinfo{year}{1959}).

\bibitem{besseling2012oscillatory}
\bibinfo{author}{Besseling, T.} \emph{et~al.}
\newblock \bibinfo{title}{Oscillatory shear-induced 3d crystalline order in
  colloidal hard-sphere fluids}.
\newblock \emph{\bibinfo{journal}{Soft Matter}} \textbf{\bibinfo{volume}{8}},
  \bibinfo{pages}{6931--6939} (\bibinfo{year}{2012}).

\bibitem{kleinert2000theory}
\bibinfo{author}{Kleinert, H.}
\newblock \bibinfo{title}{Theory and satellite experiment for critical exponent
  $\alpha$ of $\lambda$-transition in superfluid helium}.
\newblock \emph{\bibinfo{journal}{Physics Letters A}}
  \textbf{\bibinfo{volume}{277}}, \bibinfo{pages}{205--211}
  (\bibinfo{year}{2000}).

\bibitem{brady1982correction}
\bibinfo{author}{Brady, R.}
\newblock \bibinfo{title}{Correction to the formula for the london moment of a
  rotating superconductor}.
\newblock \emph{\bibinfo{journal}{Journal of Low Temperature Physics}}
  \textbf{\bibinfo{volume}{49}}, \bibinfo{pages}{1--17} (\bibinfo{year}{1982}).

\end{thebibliography}


\ifnature

\begin{addendum}

	\item We thank Ross Anderson,
	Sebastien Balibar, Robin Ball, Michael Cates, Yorgos Katsikis, Daphne Klotsa, Anthony Leggett
	and Peter Moroshkin for helpful comments
	and materials.
	
	\item[Author contributions] 
	The authors' principal contributions were: $ \gamma $ helium,
	numerical simulations and flow paradigm, R.M.B.;  
	colloids and mesophases, E.T.S.; thermodynamic calculations, D.H.P.T.
	All authors commented on and approved all aspects of the paper.

	\ifnaturesubmit
	
	\item[Competing financial interests]
	The authors declare that they have no
	competing financial interests.
	
	\item[Materials and correspondence]
	Correspondence and requests for materials
	should be addressed to E.T.S. ~(email: et@unc.edu).

	\fi
\end{addendum}

\else
\bibliographystyle{unsrt}
\textbf{Acknowledgements} We thank 
Sebastien Balibar, 
Robin Ball, 
Michael Cates, 
Peter Moroshkin 
and 
David Turban
for helpful information and comments.
\fi  

\fi

\end{document}